\documentclass[prd,preprintnumbers,nofootinbib]{revtex4} 
\usepackage{graphicx} 
\usepackage{amsmath}
\usepackage{amsfonts,amsbsy}
\usepackage{amssymb}
\usepackage{appendix}

\def\zb{{\bar z}}
\def\bz{{\bar {z}}}

\def\beq{\begin{equation}}
\def\eeq{\end{equation}}
\def\beq{\begin{eqnarray}}
\def\eeq{\end{eqnarray}}

\def\tr{{\rm tr}\,}

\begin{document}

\title{\bf Double and triple inclusive gluon production at mid rapidity: quantum interference in p-A scattering}

\author{Tolga Altinoluk$^a$, N\'{e}stor Armesto$^b$, Alex Kovner$^{c,d}$, Michael Lublinsky$^e$}

\affiliation{$^a$National Centre for Nuclear Research, 00-681 Warsaw, Poland\\
$^b$Departamento de F\'{i}sica de Part\'{i}culas and IGFAE, \\Universidade de Santiago de Compostela, 15782 Santiago de Compostela, Galicia-Spain\\
$^c$Physics Department, University of Connecticut, 2152 Hillside road, Storrs, CT 06269, USA\\
$^d$Theoretical Physics Department, CERN, CH-1211 Geneve 23, Switzerland\\
$^e$Physics Department, Ben-Gurion University of the Negev, Beer Sheva 84105, Israel
}


\date{\today}

\preprint{CERN-TH-2018-118}

\begin{abstract}
We compute double and triple inclusive gluon production in p-A scattering beyond the so-called ``glasma graph'' approximation. We consider quantum interference effects and identify in this general setup the terms responsible for the gluon HBT and initial wave function Bose enhancement which lead to correlations in particle production. Both of these terms originate from the factorizable part of the quadrupole and sextupole terms in the  production cross section. We also show that the target Bose enhancement in this regime is suppressed at large number of colors.
\end{abstract}

\maketitle

\section{Introduction}
\label{intro}
The interest in correlated particle production in the recent years has been triggered by the observation at the Large Hadron Collider (LHC) of the so called ridge correlations in p-p and p-Pb scattering~\cite{ridge}.  Many features of the ridge correlations are shared in these reactions with similar observations in Pb-Pb collisions at the LHC and earlier observations of the same structure in Au-Au collisions at the Relativistic Heavy Ion Collider, later observed in d-Au and $^3$He-Au \cite{ridge2}.
In heavy ion collisions the accepted explanation of the origin of correlations is due to a collective behavior of the final state of the dense system of gluons, which follows a hydrodynamic evolution starting a short time after scattering. A similar explanation has been put forward for the observed ridge in p-A and p-p  as well~\cite{bozek}.
 Nevertheless, the question remains of whether the mechanism that leads to these correlations in small systems (p-p and p-A)  is of the same origin.
 
 It has been suggested that the structure of the wave function of the highly energetic proton can be nontrivial and contain preexisting correlations which are reflected in the final state of the scattering~\cite{gg}. Motivated by this idea, a reasonable description of large parts of the data has been provided by calculations based on the Color Glass Condensate (CGC) approach in \cite{kevin}. This work used the so called ``glasma graphs'' approximation which is based essentially on the dilute-dilute limit of the CGC framework \cite{gg}.
 
   The physics of ``glasma graph'' approximation used in \cite{kevin} was elucidated later in \cite{us}. There it was shown that the physical origin of the correlations in the ``glasma graph'' approach where the quantum interference effects that unavoidably appear in a system of identical bosons, i.e. gluons. The two distinct contributing quantum effects identified in \cite{us} were the Bose enhancement of gluons in the incoming projectile wave function and the Hanbury Brown-Twiss (HBT)  interference effect in the emission of gluons in the final state.  Diagrammatics of the HBT effect  in QCD was discussed in \cite{earlyhbt}, and in the CGC framework more recently in \cite{yuri}. Similar effects (albeit with opposite sign) have been later identified in the quark production in the CGC based framework in \cite{PB}, see also \cite{Martinez:2018ygo,Martinez:2018tuf}. The glasma graph approach was also used in \cite{Ozonder:2014sra,Ozonder:2017wmh} to study triple and quadruple gluon correlations. However, these studies were performed in the dilute-dilute limit of glasma graphs which should be only applicable for p-p collisions at midrapidity, while we aim to go beyond this limit and consider p-A. 

We mention another recent paper \cite{urs} where quantum interference effects for gluons were studied in a model calculation from the point of view of multi parton interactions. More recently a calculation of quantum interference effects in forward quark production in the framework of multi parton interactions has been performed in \cite{amir}. There it was shown that these effects are ubiquitous in identical particle production, and that in the framework of the eikonal multiple scattering approach are contained in the quadrupole amplitude.
  
 It is desirable to have a similar detailed understanding of quantum interference effects in  gluon production at mid rapidity in the framework of the unabridged CGC formulation, i.e., beyond the ``glasma graph'' approximation\footnote{ Note that several other attempts exist in the literature to go beyond glasma graphs in multi particle production \cite{dipolesensembles}. These works however do not include the quantum interference effects, but rather deal with a more careful evaluation of non factorizable contributions to products of dipoles within the McLerran-Venugopalan model \cite{mv}. There is also a body of literature devoted to the evaluation of higher multipoles within the same framework \cite{multipoles}, as well as discussing their evolution with energy, however the relevance of these objects to multi gluon production has not been elucidated.}.
 
  The purpose of the present paper is precisely to provide such a calculation for two and three particle inclusive production. We perform it within the dilute-dense CGC approach which is a fully consistent scheme for calculation of particle production in p-A collisions.  
  In this respect we follow the same line as \cite{aklm} and \cite{yuri}. We generalize the recent work along the same lines \cite{Altinoluk:2018hcu} (which is based on the approach of  \cite{yuri}) by identifying more generally the origin of the various terms in the double inclusive gluon production as well as calculating quantum interference effects in three gluon production for the first time.
  
 In order to identify various physical effects, in the present paper we adapt the approach of \cite{amir} to production of particles in the adjoint representation.  We use the insight of \cite{yuri,amir} regarding the approximation of the quadrupole amplitude in terms of dipoles and explain similarly to \cite{amir} that this approximation correctly accounts for the quantum interference effecs. This considerably simplifies the color algebra, which in general leads to several different Wilson line ensembles \cite{yuri,Nikolaev:2005zj} whose target averages have to be modelled. It also allows us to identify in full generality the Bose enhancement (both in the projectile and the target wave function) as well as HBT contributions to particle production within this general framework, besides providing results beyond any approximation of large number of colors $N_c$.
  
  We show that just like for fundamentally charged particles, the effects due to the quantum statistics that are encoded in the quadrupole amplitude are leading correlation effects at large $N_c$. They contribute to the correlated production at order $1/N_c^2$, while similar terms that arise from the dipole squared (no color exchange) term are suppressed as $1/N_c^4$.
  We also observe that in the dilute-dense framework the terms in the production amplitude that arise due to Bose enhancement in the target are suppressed by $1/N_c^2$ relative to the Bose enhancement terms in the projectile. This $N_c$ counting is quite different from that in the glasma graph approximation, where both effects are of the same order in $1/N_c$.
 
The plan of this paper is the  following. In Section \ref{section:Double_Inc} we perform the calculation of the double inclusive gluon production at central rapidities in p-A. We describe the formalism and set out the approximation that we are using for calculating the target averages for a saturated target  with confinement radius given by the inverse of saturation momentum. Besides, we discuss the meaning of various contributions to correlated production, identifying the Bose enhancement and HBT ones. We also show that the interesting correlated terms in this expression arise from the contribution of the quadrupole in the cross section. Section \ref{section:Triple_Inc} contains the calculation for the triple inclusive gluon production. In Section IV we provide a short discussion of our results.

\section{The double inclusive gluon production in dilute-dense scattering}
\label{section:Double_Inc}

\begin{figure}[hbt]
\begin{center}
\vspace*{0.5cm}
\includegraphics[width=10cm]{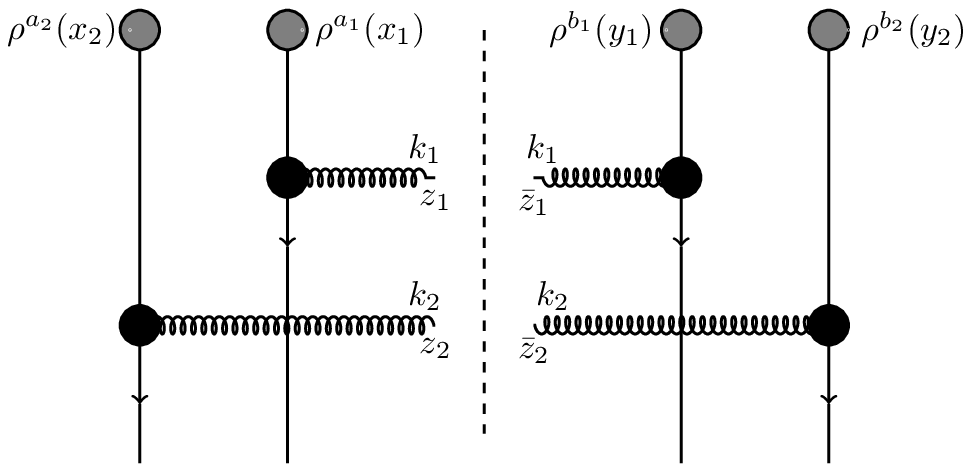}
\end{center}
\caption{Graphical illustration of Eq. \eqref{X_section_formal} with the vertical lines representing the rescatterings with the target through Wilson lines.}
\label{fig1}
\end{figure}

The aim of this Section is to calculate the double inclusive gluon production in p-A collisions in terms of the dipole scattering amplitude. Our starting point is the general expression \cite{double} for production of two gluons with pseudorapidities $\eta_1, \eta_2$ and transverse momenta $k_1,k_2$
\beq
\label{X_section_formal}
\frac{d\sigma}{d^2k_1d\eta_1d^2k_2d\eta_2}&=&\alpha_s^2(4\pi)^2\int_{z_1\bz_1z_2\bz_2} e^{ik_1\cdot(z_1-\bz_1)+ik_2\cdot(z_2-\bz_2)}\nonumber \\
&\times& \int_{x_1x_2y_1y_2}A^i(x_1-z_1)A^i(\bz_1-y_1)A^j(x_2-z_2)A^j(\bz_2-y_2)
\Big\langle \rho^{a_1}(x_1)\rho^{a_2}(x_2)\rho^{b_1}(y_1)\rho^{b_2}(y_2)\Big\rangle_P\nonumber\\
&\times&
\Big\langle \big[U(z_1)-U(x_1) \big]^{a_1c} \big[ U^\dagger(\bz_1)-U^\dagger(y_1)\big]^{cb_1}\big[ U(z_2)-U(x_2) \big]^{a_2d} \big[ U^\dagger(\bz_2)-U^\dagger(y_2)\big]^{db_2}\Big\rangle_T \,.
\eeq
Here, $\rho^a(x)$ is the color charge density in the projectile,  $U(x)$ is the adjoint Wilson line in the color field of the target representing the scattering matrix of a gluon at  transverse coordinate $x$, $a$ is the color index running from $1$ to $N_c^2-1$, $\int_z\equiv \int d^2z$ and the Weiszacker-Williams field $A^i$ is given by
\beq
\label{A_in_momentum_space}
A^i(x-y)=-\frac{1}{2\pi}\frac{(x-y)_i}{(x-y)^2}=\int \frac{d^2k}{(2\pi)^2}e^{-ik\cdot(x-y)}\frac{k^i}{k^2}\,.
\eeq
Eq. \eqref{X_section_formal} is graphically illustrated in Fig. \ref{fig1}.

The averaging over $\rho$, $\langle \cdots \rangle_P$,  in Eq. \eqref{X_section_formal} will be performed using the McLerran-Venugopalan  (MV) model \cite{mv}.
As for averaging over $U(x)$, $\langle \cdots \rangle_T$, it will not be performed explicitly but some properties of the distribution of $U$'s will be used to simplify and interpret our general expressions. 

\subsection{Projectile averaging in double inclusive gluon production}
\label{doubleprojave}
We start with averaging of the cross section with respect to the projectile color charge distribution. We use the generalized MV model where the weight functional is Gaussian. Thus the average of any product of the color charge densities factorizes into a product of all possible Wick contractions:
\beq
\Big\langle \rho^{a_1}(x_1) \rho^{a_2}(x_2) \rho^{b_1}(y_1) \rho^{b_2}(y_2) \Big\rangle_P=
\Big\langle \rho^{a_1}(x_1) \rho^{a_2}(x_2) \Big\rangle_P
\Big\langle \rho^{b_1}(y_1) \rho^{b_2}(y_2) \Big\rangle_P\nonumber\\
+\,
\Big\langle \rho^{a_1}(x_1) \rho^{b_1}(y_1) \Big\rangle_P
\Big\langle \rho^{a_2}(x_2) \rho^{b_2}(y_2) \Big\rangle_P\nonumber\\
+\,
\Big\langle \rho^{a_1}(x_1) \rho^{b_2}(y_2) \Big\rangle_P
\Big\langle \rho^{a_2}(x_2) \rho^{b_1}(y_1) \Big\rangle_P.
\eeq 
For the average of two projectile color charges we take a general form:
\beq
\label{rho_rho}
\Big\langle \rho^{a}(x) \rho^{b}(y) \Big\rangle_P=\delta^{ab} \mu^2(x,y).
\eeq
We do not assume translational invariance of the projectile wave function. This means that the function $\mu^2(x,y)$ depends both on the difference $x-y$ and the center-of-mass coordinate $\frac{x+y}{2}$. The finite transverse size of the projectile $R$ is reflected in vanishing of $\mu^2(x,y)$ for $(x+y)^2>4R^2$.

Then the average of four projectile color charges reads
\beq
\label{Av_4_rho}
\Big\langle \rho^{a_1}(x_1) \rho^{a_2}(x_2) \rho^{b_1}(y_1) \rho^{b_2}(y_2) \Big\rangle_P=
\delta^{a_1a_2}\delta^{b_1b_2} \,  \, \mu^2(x_1,x_2) \, \mu^2(y_1,y_2)
+
\delta^{a_1b_1}\delta^{a_2b_2} \, \, \mu^2(x_1,y_1) \, \mu^2(x_2,y_2)
\nonumber\\
+
\delta^{a_1b_2}\delta^{a_2b_1} \, \, \mu^2(x_1,y_2) \, \mu^2(x_2,y_1).
\eeq

Implementing this projectile averaging procedure we get
\beq
\label{X_section_without_target_pairwise_contraction}
\frac{d\sigma}{d^2k_1d\eta_1d^2k_2d\eta_2}=\alpha_s^2(4\pi)^2\int_{z_1\bz_1z_2\bz_2} e^{ik_1\cdot(z_1-\bz_1)+ik_2\cdot(z_2-\bz_2)}\int_{x_1x_2y_1y_2}A^i(x_1-z_1)A^i(\bz_1-y_1)A^j(x_2-z_2)A^j(\bz_2-y_2)
\nonumber\\
\times 
\; 
 \; 
\Bigg\{ 
\mu^2(x_1,x_2) \, \mu^2(y_1,y_2)\; 
\bigg\langle \tr \Big\{
\big[U(z_1)-U(x_1) \big]\big[ U^\dagger(\bz_1)-U^\dagger(y_1)\big]
\big[ U(\bz_2)-U(y_2)\big]\big[ U^\dagger(z_2)-U^\dagger(x_2) \big] \Big\}
\bigg\rangle_T
\nonumber\\
+
\mu^2(x_1,y_1) \, \mu^2(x_2,y_2)\;
\bigg\langle 
\tr\Big\{ \big[U(z_1)-U(x_1) \big] \big[ U^\dagger(\bz_1)-U^\dagger(y_1)\big]\Big\}
\tr\Big\{  \big[ U(z_2)-U(x_2) \big] \big[ U^\dagger(\bz_2)-U^\dagger(y_2)\big] \Big\}\bigg\rangle_T
\nonumber
\\
+
\mu^2(x_1,y_2) \, \mu^2(x_2,y_1)
\bigg\langle
\tr\Big\{
\big[U(z_1)-U(x_1) \big] \big[ U^\dagger(\bz_1)-U^\dagger(y_1)\big]
\big[ U(z_2)-U(x_2) \big] \big[ U^\dagger(\bz_2)-U^\dagger(y_2)\big] \Big\}\bigg\rangle_T\Bigg\}.
\eeq
We define the dipole and the quadrupole  amplitudes in the standard way as 
\beq
\label{dipole_def}
D(x,y)&=&\frac{1}{N_c^2-1}\tr\big[ U(x)U^\dagger(y)\big],\\
\label{Q_def}
Q(x,y,z,v)&=&\frac{1}{N_c^2-1}\tr \big[ U(x)U^\dagger(y) U(z)U^\dagger(v) \big],
\eeq
and the corresponding Fourier transforms as
\beq
\label{Trans_Inv_dipole}
D(x_1,x_2)&=&\int \frac{d^2q_1}{(2\pi)^2} \frac{d^2q_2}{(2\pi)^2}e^{-iq_1\cdot x_1+ i q_2 \cdot x_2} D(q_1,q_2)\ ,\\
\label{Trans_Inv_Q}
Q(x_1,x_2,x_3,x_4)&=&\int \frac{d^2q_1}{(2\pi)^2}\frac{d^2q_2}{(2\pi)^2}\frac{d^2q_3}{(2\pi)^2}\frac{d^2q_4}{(2\pi)^2}
e^{-iq_1\cdot x_1+ i q_2 \cdot x_2-iq_3\cdot x_3+ i q_4 \cdot x_4} Q(q_1,q_2,q_3,q_4)\ .
\eeq
The cross section is then written most conveniently as the sum of three terms:
\beq
\label{Final_Type_A_Q}
&&
\frac{d\sigma}{d^2k_1d\eta_1d^2k_2d\eta_2}\bigg|_{\rm type\; A}=\alpha_s^2(4\pi)^2 (N_c^2-1)
\int
\frac{d^2q_1}{(2\pi)^2}\frac{d^2q_2}{(2\pi)^2}\frac{d^2q_3}{(2\pi)^2}\frac{d^2q_4}{(2\pi)^2} \; \langle Q(q_1,q_2,q_3,q_4) \rangle_T
\nonumber\\
&&\times \; 
\mu^2\Big[ (k_1-q_1), (k_2+q_4) \Big] 
\mu^2 \Big[ -(k_1-q_2), -(k_2+q_3) \Big]\; L^i(k_1,q_1)L^i(k_1,q_2) \, L^j(k_2,-q_3)L^j(k_2,-q_4),
\label{typeA}
\eeq
\beq
\label{Final_Type_B_dd}
&&
\frac{d\sigma}{d^2k_1d\eta_1d^2k_2d\eta_2}\bigg|_{\rm type\; B}=\alpha_s^2(4\pi)^2 (N_c^2-1)^2
\int
\frac{d^2q_1}{(2\pi)^2}\frac{d^2q_2}{(2\pi)^2}\frac{d^2q_3}{(2\pi)^2}\frac{d^2q_4}{(2\pi)^2}\langle D(q_1,q_2)D(q_3,q_4)\rangle_T
\\
&&
\times \; 
\mu^2\Big[ (k_1-q_1), -(k_1-q_2) \Big] 
\mu^2 \Big[ (k_2-q_3), -(k_2-q_4) \Big]\nonumber
L^i(k_1,q_1)L^i(k_1,q_2)\; L^j(k_2,q_3)L^j(k_2,q_4)
\label{typeB}
\eeq
and 
\beq
\label{Final_Type_C_Q}
&&
\frac{d\sigma}{d^2k_1d\eta_1d^2k_2d\eta_2}\bigg|_{\rm type\; C}=\alpha_s^2(4\pi)^2 (N_c^2-1)
\int
\frac{d^2q_1}{(2\pi)^2}\frac{d^2q_2}{(2\pi)^2}\frac{d^2q_3}{(2\pi)^2}\frac{d^2q_4}{(2\pi)^2} \; \langle Q(q_1,q_2,q_3,q_4)\rangle_T
\nonumber\\
&&
\times \; 
\mu^2\Big[ (k_1-q_1), -(k_2-q_4) \Big] 
\mu^2 \Big[ (k_2-q_3), -(k_1-q_2) \Big]\,
L^i(k_1,q_1)L^i(k_1,q_2) \, L^j(k_2,q_3)L^j(k_2,q_4),
\label{typeC}
\eeq
where, for convenience, we have defined the Lipatov vertex
\begin{equation}
\label{L_def}
L^i(k,q)\equiv \frac{(k-q)^i}{(k-q)^2}-\frac{k^i}{k^2}\ .
\end{equation}

\begin{figure}[hbt]
\begin{center}
\vspace*{0.5cm}
\includegraphics[width=9cm]{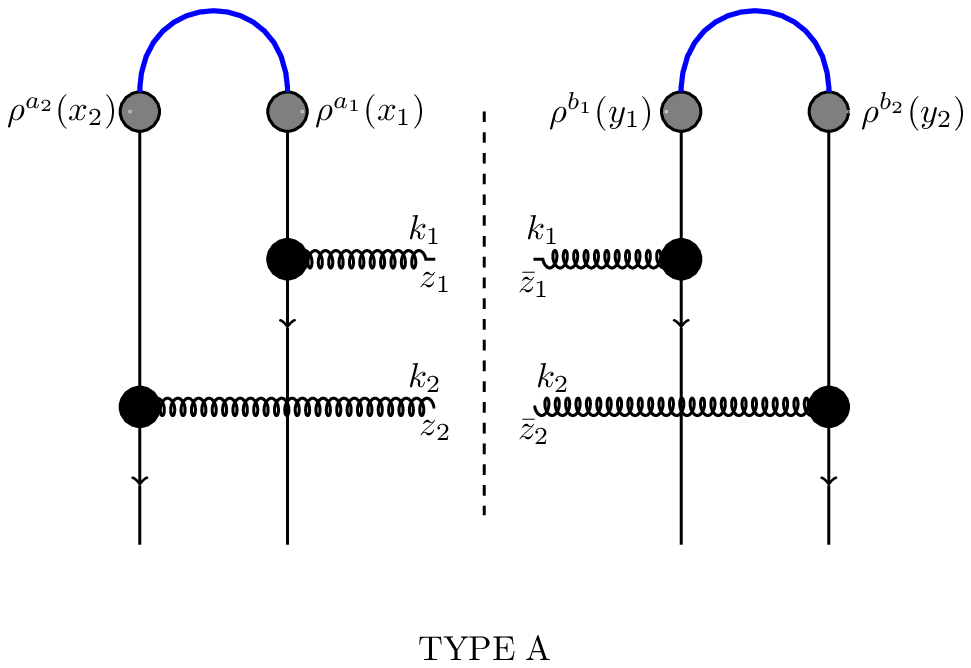}
\end{center}
\caption{Diagrams of type A, Eq. \eqref{typeA}.}
\label{fig2}
\end{figure}

\begin{figure}[hbt]
\begin{center}
\vspace*{0.5cm}
\includegraphics[width=9cm]{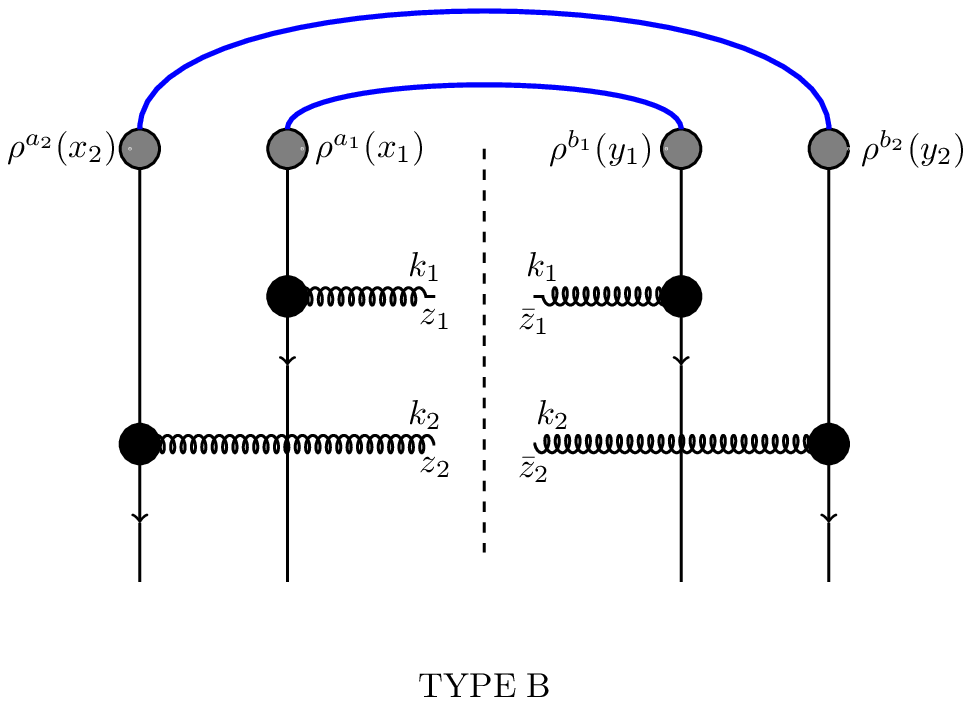}
\end{center}
\caption{Diagrams of type B, Eq. \eqref{typeB}.}
\label{fig3}
\end{figure}

\begin{figure}[hbt]
\begin{center}
\vspace*{0.5cm}
\includegraphics[width=9cm]{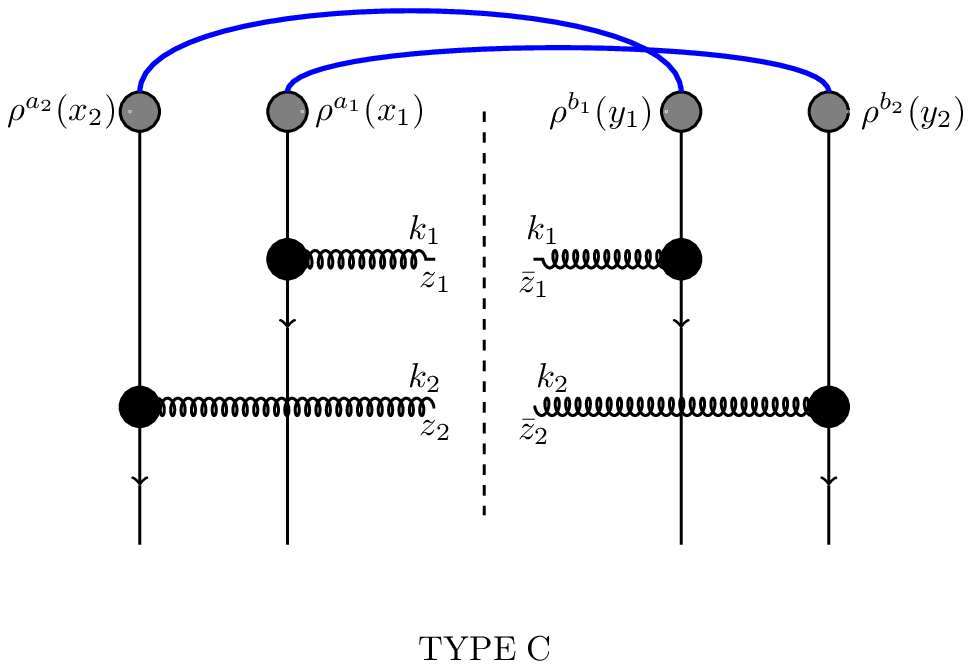}
\end{center}
\caption{Diagrams of type C, Eq. \eqref{typeC}.}
\label{fig4}
\end{figure}

The different contractions of the projectile color charge density leading to the three terms are illustrated in Figs. \ref{fig2}, \ref{fig3} and \ref{fig4}.

\subsection{Target averaging in double inclusive gluon production}
\label{doubletarave}
Our next step is to understand the general features of the target averages. Here we follow the logic of \cite{amir}. 

The cross section involves integration over the four coordinates of the product of the eikonal matrices $U(x)$. One therefore expects that the main contribution will come from the region of the transverse plain where as many points are far away from each other as possible. Configurations in which points come close to each other in the transverse plain will give contributions suppressed by powers of area of the projectile.

On the other hand, one cannot have all four points far away from each other. This follows since the target field ensemble has to be color invariant. As a mater of fact, it is reasonable to assume that the color neutralization in the target ensemble is achieved on transverse distance scales of order $1/Q_s$. In order for the $S$-matrix on such a target to be non vanishing, the objects that scatter must be color singlets of size of order or smaller than $1/Q_s$. Thus, the maximal contribution must come from the configurations where the four points are combined into pairs, such that each pair is a singlet and the distance between the pairs is large. Taking into account only such configurations is equivalent to the calculation of target averages in which one factorizes the average of a product of any number of $U$ matrices into averages of pairs with the basic ``Wick contraction'' given by 
\beq
\label{Average_2W}
\Big\langle U^{ab}(x)U^{cd}(y)\Big\rangle_T= \delta^{ac}\delta^{bd}\frac{1}{(N_c^2-1)^2}\Big\langle \tr[U(x)U^\dagger(y)] \Big\rangle_T=\delta^{ac}\delta^{bd}\frac{1}{N_c^2-1}d(x,y),
\eeq
where
\begin{equation}
d(x,y)\equiv \langle D(x,y)\rangle_T.
\end{equation}
Note that only one color structure appears in this expression, the one where the left and right indexes of the two $U$-matrices are in color singlets. The physical reason for this is the following. Recall that the left index of $U$ specifies the color of the incoming gluon, while the right index  the color of the scattered gluon. We expect that on a saturated target the $S$-matrix of any nonsinglet colored state  vanishes (black disk limit). The structure of Eq. \eqref{Average_2W} encodes precisely this property.

With these physical assumptions on the target field ensemble we have
\begin{eqnarray}
\langle Q(x,y,z,v)\rangle_T&\longrightarrow&d(x,y)d(z,v)+d(x,v)d(z,y)+\frac{1}{N_c^2-1}d(x,z)d(y,v),\\
\langle D(x,y)D(z,v)\rangle_T&\longrightarrow&d(x,y)d(z,v)+\frac{1}{(N_c^2-1)^2}\left[d(x,v)d(y,z)+d(x,z)d(v,y)\right].
\end{eqnarray}
It is now straightforward to rewrite the double gluon inclusive production cross section entirely in terms of the dipole averages
assuming translational invariance,
\beq
\label{Trans_Inv_dipole2}
d(x_1,x_2)&=&\int \frac{d^2q_1}{(2\pi)^2} \frac{d^2q_2}{(2\pi)^2}e^{-iq_1\cdot x_1+ i q_2 \cdot x_2} \ d\left(\frac{q_1+q_2}{2}\right)\delta^{(2)}(q_1-q_2)\ .
\eeq
We find

\beq
\label{X_section_Type_A}
&&
\frac{d\sigma}{d^2k_1d\eta_1d^2k_2d\eta_2}\bigg|_{\rm type \; A}=\alpha_s^2(4\pi)^2 
\int \frac{d^2q_1}{(2\pi)^2} \frac{d^2q_2}{(2\pi)^2} d(q_1) d(q_2)\\
&&\times
\Big\{
(N_c^2-1)\; 
\mu^2\Big[ (k_1-q_1),(k_2-q_2)\Big]\mu^2\Big[ (q_1-k_1),(q_2-k_2)\Big]
L^i(k_1,q_1)L^i(k_1,q_1)\; L^j(k_2,q_2)L^j(k_2,q_2)
\nonumber\\
&&+\; 
(N_c^2-1)\; 
\mu^2\Big[ (k_1-q_1), (k_2+q_1)\Big]\mu^2\Big[ -(k_1-q_2), -(q_2+k_2)\Big]
L^i(k_1,q_1)L^i(k_1,q_2)\; L^j(k_2,-q_1)L^j(k_2,-q_2) 
\nonumber\\
&&+
\; 
\mu^2\Big[ (k_1-q_1),(k_2-q_2)\Big]\mu^2\Big[-(k_1-q_2),(q_1-k_2)\Big] 
L^i(k_1,q_1)L^i(k_1,q_2)\; L^j(k_2,q_1)L^j(k_2,q_2)\Big\}, \nonumber
\eeq
\beq
\label{X_section_Type_B}
&&
\frac{d\sigma}{d^2k_1d\eta_1d^2k_2d\eta_2}\bigg|_{\rm type\; B}=\alpha_s^2(4\pi)^2
  \, 
\int \frac{d^2q_1}{(2\pi)^2} \frac{d^2q_2}{(2\pi)^2} d(q_1) d(q_2)\\
&&\times
\Big\{
(N_c^2-1)^2\; 
\mu^2\Big[(k_1-q_1), (q_1-k_1)\Big]\mu^2\Big[(k_2-q_2),(q_2-k_2)\Big]
L^i(k_1,q_1)L^i(k_1,q_1)\; L^j(k_2,q_2)L^j(k_2,q_2)
\nonumber\\
&&+\;
\mu^2\Big[ (k_1-q_1), -(k_1-q_2) \Big]\mu^2\Big[ (k_2+q_1),-(q_2+k_2)\Big] L^i(k_1,q_1)L^i(k_1,q_2)\; L^j(k_2,-q_1)L^j(k_2,-q_2)
\nonumber\\
&&+\; 
\mu^2\Big[ (k_1-q_1), -(k_1-q_2)\Big]\mu^2\Big[ (k_2-q_2), (q_1-k_2)\Big] L^i(k_1,q_1)L^i(k_1,q_2)\; L^j(k_2,q_1)L^j(k_2,q_2)\Big\},
\nonumber
\eeq
\beq
\label{X_section_Type_C}
&&
\frac{d\sigma}{d^2k_1d\eta_1d^2k_2d\eta_2}\bigg|_{\rm type\; C}=\alpha_s^2(4\pi)^2 
 \,
\int \frac{d^2q_1}{(2\pi)^2} \frac{d^2q_2}{(2\pi)^2} d(q_1) d(q_2)\\
&&\times
\Big\{
(N_c^2-1)\; \mu^2\Big[ (k_1-q_1), (q_2-k_2)\Big]\mu^2\Big[ (k_2-q_2), (q_1-k_1)\Big]
L^i(k_1,q_1)L^i(k_1,q_1)\; L^j(k_2,q_2)L^j(k_2,q_2)
\nonumber\\
&&+\; 
\mu^2\Big[(k_1-q_1), (q_2-k_2)\Big]\mu^2\Big[ (k_2+q_1), -(k_1+q_2)\Big]
L^i(k_1,q_1)L^i(k_1,-q_2)\; L^j(k_2,-q_1)L^j(k_2,q_2)
\nonumber\\
&&+\;(N_c^2-1)\; 
\mu^2\Big[ (k_1-q_1), (q_1-k_2)\Big]\mu^2\Big[(k_2-q_2), -(k_1-q_2)\Big]
L^i(k_1,q_1)L^i(k_1,q_2)\; L^j(k_2,q_1)L^j(k_2,q_2)\Big\}.
\nonumber\\
\nonumber
\eeq

Finally let us organize the terms in powers of $N_c^2-1$. Then
\begin{equation}
\frac{d\sigma}{d^2k_1d\eta_1d^2k_2d\eta_2}=\alpha_s^2(4\pi)^2 (N_c^2-1)^2
\int \frac{d^2q_1}{(2\pi)^2} \frac{d^2q_2}{(2\pi)^2} d(q_1) d(q_2)
\Bigg\{I_0+\frac{1}{N_c^2-1}I_1+\frac{1}{(N_c^2-1)^2}I_2\Bigg\},
\label{doubleres}
\end{equation}
with
\beq\label{is}
I_0&=&\mu^2\Big[(k_1-q_1), (q_1-k_1)\Big]\mu^2\Big[(k_2-q_2),(q_2-k_2)\Big]L^i(k_1,q_1)L^i(k_1,q_1)L^j(k_2,q_2)L^j(k_2,q_2),\\
I_1&=& \mu^2\Big[ (k_1-q_1), (q_2-k_2)\Big]\mu^2\Big[ (k_2-q_2), (q_1-k_1)\Big]L^i(k_1,q_1)L^i(k_1,q_1)L^j(k_2,q_2)L^j(k_2,q_2)\\
&+&\mu^2\Big[ (k_1-q_1), (q_1-k_2)\Big]\mu^2\Big[(k_2-q_2), -(k_1-q_2)\Big]L^i(k_1,q_1)L^i(k_1,q_2)L^j(k_2,q_1)L^j(k_2,q_2)\nonumber\\
&+&(k_2\rightarrow -k_2),\nonumber\\
I_2&=&\mu^2\Big[ (k_1-q_1), -(k_1-q_2)\Big]\mu^2\Big[ (k_2-q_2), (q_1-k_2)\Big]L^i(k_1,q_1)L^i(k_1,q_2)L^j(k_2,-q_1)L^j(k_2,-q_2)+(k_2\rightarrow -k_2)\\
&+&\mu^2\Big[ (k_1-q_1),-(k_2-q_2)\Big]\mu^2\Big[-(k_1+q_2),(q_1+k_2)\Big]L^i(k_1,q_1)L^i(k_1,-q_2)L^j(k_2,-q_1)L^j(k_2,q_2)+(k_2\rightarrow -k_2).\nonumber
\eeq

\subsection{Identifying  terms in double inclusive gluon production}
\label{doubleiden}
Given the expressions above it is quite straightforward to identify the physical origin of the various terms above.

First of all we note, that the cross section is symmetric under the transformation $(k_1,k_2)\rightarrow (k_1,-k_2)$. This property of the dilute-dense approximation is well known in the literature. It is also known that the symmetry is ``accidental'' and it disappears once one includes higher order perturbative corrections \cite{v3}. We will therefore only consider half of the terms 
in Eq. \eqref{is}, namely those that are written out explicitly.

To understand the meaning of the various terms we have to first of all specify $\mu^2(k,p)$ a little further. Recall that it is defined as thee correlation function of the color charge density, which operationally reads
\begin{equation}
\mu^2(k,p)=\frac{1}{N_c^2-1}\langle \rho^a(k)\rho^a(p)\rangle_P \ .
\end{equation}
In the hypothetical translational invariant limit, where the distribution of $\rho$ is invariant under translations in the transverse plain, we would have
\begin{equation}
\mu^2_{TI}(k,p)=T\left(\frac{k-p}{2}\right)\delta^{(2)}(k+p).
\end{equation}
The translationally invariant approximation is quite reasonable if we are interested in production of high transverse momentum gluons, however in a more accurate calculation the transverse size of the projectile should be reflected in $\mu^2$. A reasonable way to include it is to introduce a form factor of the type
\begin{equation}
\mu^2(k,p)=T\left(\frac{k-p}{2}\right)F[(k+p)R],
\end{equation}
where $R$ is the radius of the projectile.
Here $T$ is roughly speaking the transverse dependent distribution of the valence charges, and $F(x)$ is a soft form factor which  is maximal at $x=0$ and rapidly decreases  to zero at $|x|>1$.  The exact form of the function $F(x)$ does not matter for our purposes, it can be taken as a Gaussian $F_G(x)=\exp{(-x^2)}$ , or as a Lorentzian $F_L(x)=1/(1+x^2)$, or any other function with these properties. The important property is that $F(x)$ vanishes when the sum of the transverse momenta is not soft.

Let us now examine the various terms in Eq. \eqref{is}:
\begin{itemize}
\item{First off, the term $I_0$ obviously is just an uncorrelated production cross section, which is equal to the square of single gluon emission probability. It is not interesting from the point of view of correlated production.}

\item{The expression for $I_1$ contains two distinct terms, and it is easy to see that they have quite different origin. The first term 
is proportional to 
\begin{equation}
\mu^2\Big[ (k_1-q_1), (q_2-k_2)\Big]\mu^2\Big[ (k_2-q_2), (q_1-k_1)\Big]=T^2\left(\frac{k_1+k_2-q_1-q_2}{2}\right)F^2\left[\Big(k_1-q_1-(k_2-q_2)\Big)R\right].
\end{equation}
Note that the momenta $k_1-q_1$ and $k_2-q_2$ are the momenta the two gluons have in the projectile wave function, since $k_1$ and $k_2$ are the momenta in the final state, while $q_1$ and $q_2$, being the arguments of the dipole scattering amplitudes, are the momentum transfers imparted to the two gluons during the propagation through the target.  
Due to the properties of the form factor $F$, this term is sharply peaked when the momenta of the two gluons in the projectile wave function are very close to each other, i.e., within the inverse projectile radius. This term therefore embodies the Bose enhancement in the incoming projectile wave function.}

\item{The nature of the second term in $I_1$ in Eq. \eqref{is} is defined by the factor
\begin{equation} 
\mu^2\Big[ (k_1-q_1), (q_1-k_2)\Big]\mu^2\Big[(k_2-q_2), -(k_1-q_2)\Big]=T\left(\frac{k_1+k_2}{2}-q_1\right)T\left(\frac{k_1+k_2}{2}-q_2\right)F^2\left[(k_1-k_2)R\right].
\end{equation}
This term enhances production of pairs of gluons with equal (up to $1/R$) transverse momenta in the final state. This is a typical HBT contribution.}

\item{The first term in $I_2$ is proportional to
\begin{equation} 
\mu^2\Big[ (k_1-q_1), -(k_1-q_2)\Big]\mu^2\Big[(k_2-q_2), -(k_2-q_1)\Big]\propto F^2\left[(q_1-q_2)R\right].
\end{equation}
Here the momentum exchange in the scattering of two gluons is the same. Naturally this term is associated with Bose enhancement in the target wave function. This term is somewhat different from the others in that it looks like the target gluon distribution here is regulated by the projectile size $R$. This is in fact natural. The Bose enhancement in the target wave function should be certainly regulated by the target and not the projectile size. However, very long transverse wave length gluons of the target are not probed by a smaller projectile, and thus do not contribute to the cross section. This is the reason why even though the enhancement is due to the properties of the target wave function, in the expression for the cross section the regulator is the projectile size.

It is interesting to note that in the glasma graph approximation the projectile and the target Bose enhancement terms appear at the same order in $1/N_c$. On the other hand in the proper dilute-dense treatment the target Bose enhancement terms come with further suppression factors. It is indeed a well known fact that some aspects of $1/N_c$ counting are different in the dilute and dense limits \cite{Ncounting}. }
\item{Finally the second term in $I_2$ has the same structure,
\begin{equation}
\mu^2\Big[ (k_1-q_1),-(k_2-q_2)\Big]\mu^2\Big[-(k_1+q_2),(q_1+k_2)\Big]\propto F^2\left[\left( k_1-k_2-q_1+q_2 \right)R\right],
\end{equation}
as the projectile Bose enhancement term and constitutes an $1/N_c^2$ suppressed correction to this effect.}

\end{itemize}

\section{The triple inclusive gluon production in dilute-dense scattering}
\label{section:Triple_Inc}

\begin{figure}[hbt]
\begin{center}
\vspace*{0.5cm}
\includegraphics[width=10cm]{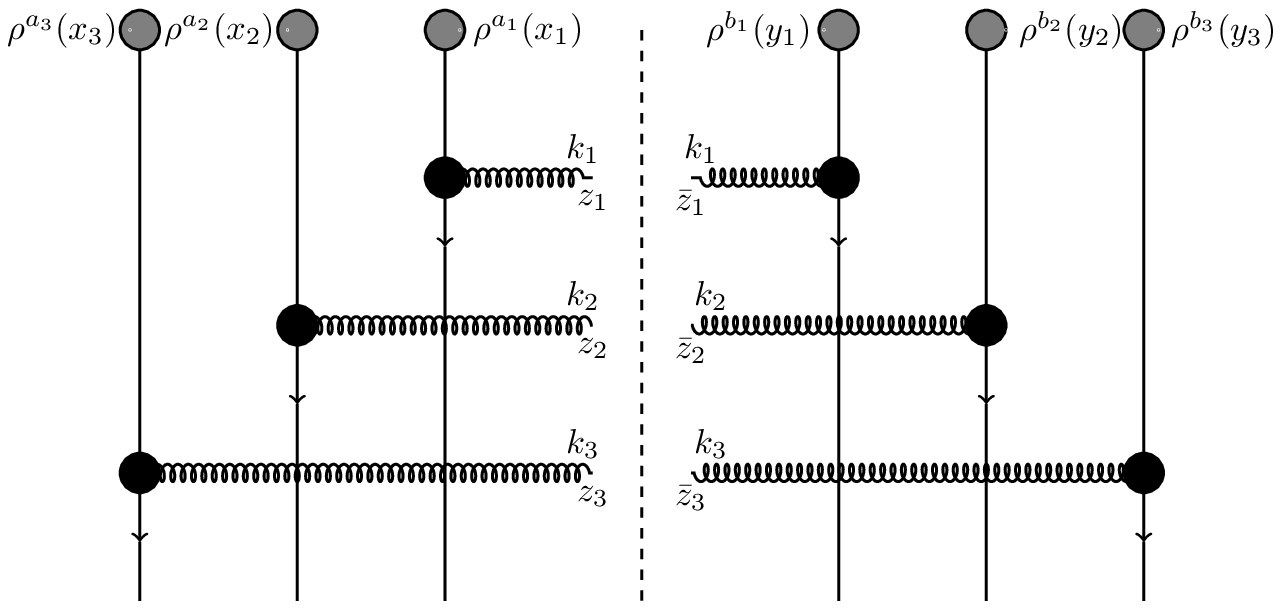}
\end{center}
\caption{Graphical illustration of Eq. \eqref{3g_Xsection} with the vertical lines representing the rescatterings with the target through Wilson lines.}
\label{fig_3g_set_up}
\end{figure}

In this section, we study inclusive triple gluon production. The set up of the process is illustrated in Fig. \ref{fig_3g_set_up}. The formal expression of the three gluon production cross section, for gluons with pseudorapidities $\eta_i$ and transverse momenta $k_i$, $i=1,2,3$, can be simply written as 
\beq
\label{3g_Xsection}
&&
\frac{d\sigma}{d^2k_1d\eta_1 \, d^2k_2d\eta_2 \, d^2k_3d\eta_3}=\alpha_s^3 \, (4\pi)^3 \, 
\int_{z_1\zb_1z_2\zb_2z_3\zb_3} e^{ik_1\cdot(z_1-\zb_1)+ik_2\cdot(z_2-\zb_2)+ik_3\cdot(z_3-\zb_3)}
 \nonumber\\
&&
\times \int _{x_1y_1 x_2y_2 x_3y_3} A^{i}(x_1-z_1)A^{i}(\zb_1-y_1)
 A^{j}(x_2-z_2)A^{j}(\zb_2-y_2) \, A^{k}(x_3-z_3)A^{k}(\zb_3-y_3) 
\nonumber\\
&&
\hspace{3cm}
\times \;  
\Big\langle \rho^{a_1}(x_1)\rho^{a_2}(x_2)\rho^{a_3}(x_3) \rho^{b_1}(y_1)\rho^{b_2}(y_2)\rho^{b_3}(y_3) \Big\rangle _P
\nonumber\\
&&
\times
\bigg\langle 
\big(U_{z_1}-U_{x_1}\big)^{a_1c_1}     \big(U^\dagger_{\zb_1}-U^\dagger_{y_1}\big)^{c_1b_1} 
\big(U_{z_2}-U_{x_2}\big)^{a_2c_2}     \big(U^\dagger_{\zb_2}-U^\dagger_{y_2}\big)^{c_2b_2}
\big(U_{z_3}-U_{x_3}\big)^{a_3c_3}     \big(U^\dagger_{\zb_3}-U^\dagger_{y_3}\big)^{c_3b_3}
\bigg\rangle_T \; ,
\eeq
where the coordinate of each Wilson line is written as a subscript. 

\subsection{Projectile averaging in triple inclusive gluon production}
\label{tripleprojave}
We first  perform the averaging over the projectile color charge densities.  As in the previous section we adopt the generalized MV model for the average of two projectile color charge densities and write down all possible Wick contractions of the product of the color charge densities. Then, the average of six projectile color charges reads
\beq
\label{proj_contraction}
&&
\Big\langle \rho^{a_1}_{x_1}\rho^{a_2}_{x_2}\rho^{a_3}_{x_3} \rho^{b_1}_{y_1}\rho^{b_2}_{y_2}\rho^{b_3}_{y_3} \Big\rangle _P=
  \big\langle \rho^{a_1}_{x_1} \rho^{b_1}_{y_1}  \big\rangle \big\langle \rho^{a_2}_{x_2} \rho^{b_2}_{y_2} \big\rangle \big\langle \rho^{a_3}_{x_3} \rho^{b_3}_{y_3} \big\rangle
+\; \big\langle \rho^{a_1}_{x_1} \rho^{b_1}_{y_1}  \big\rangle
 \Big[ \big\langle \rho^{a_2}_{x_2} \rho^{a_3}_{x_3} \big\rangle \big\langle \rho^{b_2}_{y_2} \rho^{b_3}_{y_3} \big\rangle
 + \big\langle \rho^{a_2}_{x_2} \rho^{b_3}_{y_3} \big\rangle \big\langle \rho^{a_3}_{x_3} \rho^{b_2}_{y_2} \big\rangle \Big]
\nonumber\\
&&
+ \, \big\langle \rho^{a_2}_{x_2} \rho^{b_2}_{y_2}  \big\rangle
 \Big[ \big\langle \rho^{a_1}_{x_1} \rho^{a_3}_{x_3} \big\rangle \big\langle \rho^{b_1}_{y_1} \rho^{b_3}_{y_3} \big\rangle
 + \big\langle \rho^{a_1}_{x_1} \rho^{b_3}_{y_3} \big\rangle \big\langle \rho^{a_3}_{x_3} \rho^{b_1}_{y_1} \big\rangle \Big]
 +  \, \big\langle \rho^{a_3}_{x_3} \rho^{b_3}_{y_3}  \big\rangle
 \Big[ \big\langle \rho^{a_1}_{x_1} \rho^{a_2}_{x_2} \big\rangle \big\langle \rho^{b_1}_{y_1} \rho^{b_2}_{y_2} \big\rangle
 + \big\langle \rho^{a_1}_{x_1} \rho^{b_2}_{y_2} \big\rangle \big\langle \rho^{a_2}_{x_2} \rho^{b_1}_{y_1} \big\rangle \Big]
\nonumber\\
&&
+  \, \big\langle \rho^{a_1}_{x_1} \rho^{a_2}_{x_2}  \big\rangle
 \Big[ \big\langle \rho^{a_3}_{x_3} \rho^{b_1}_{y_1} \big\rangle \big\langle \rho^{b_2}_{y_2} \rho^{b_3}_{y_3} \big\rangle
 + \big\langle \rho^{a_3}_{x_3} \rho^{b_2}_{y_2} \big\rangle \big\langle \rho^{b_1}_{y_1} \rho^{b_3}_{y_3} \big\rangle \Big]
 +  \, \big\langle \rho^{a_2}_{x_2} \rho^{a_3}_{x_3}  \big\rangle
 \Big[ \big\langle \rho^{a_1}_{x_1} \rho^{b_2}_{y_2} \big\rangle \big\langle \rho^{b_1}_{y_1} \rho^{b_3}_{y_3} \big\rangle
 + \big\langle \rho^{a_1}_{x_1} \rho^{b_3}_{y_3} \big\rangle \big\langle \rho^{b_1}_{y_1} \rho^{b_2}_{y_2} \big\rangle \Big]
\nonumber\\
&&
+  \, \big\langle \rho^{a_2}_{x_2} \rho^{b_1}_{y_1}  \big\rangle
 \Big[ \big\langle \rho^{a_1}_{x_1} \rho^{a_3}_{x_3} \big\rangle \big\langle \rho^{b_2}_{y_2} \rho^{b_3}_{y_3} \big\rangle
 + \big\langle \rho^{a_1}_{x_1} \rho^{b_3}_{y_3} \big\rangle \big\langle \rho^{a_3}_{x_3} \rho^{b_2}_{y_2} \big\rangle \Big]
 +  \, \big\langle \rho^{a_2}_{x_2} \rho^{b_3}_{y_3}  \big\rangle
 \Big[ \big\langle \rho^{a_1}_{x_1} \rho^{b_2}_{y_2} \big\rangle \big\langle \rho^{a_3}_{x_3} \rho^{b_1}_{y_1} \big\rangle
 + \big\langle \rho^{a_1}_{x_1} \rho^{a_3}_{x_3} \big\rangle \big\langle \rho^{b_1}_{y_1} \rho^{b_2}_{y_2} \big\rangle \Big] \; .
\eeq
Here, we introduce a compact notation and  write the coordinate of each color charge density as a subscript (and also omitted  the subscript $P$ from the averages).  

\begin{figure}[hbt]
\begin{center}
\vspace*{0.5cm}
\includegraphics[width=8cm]{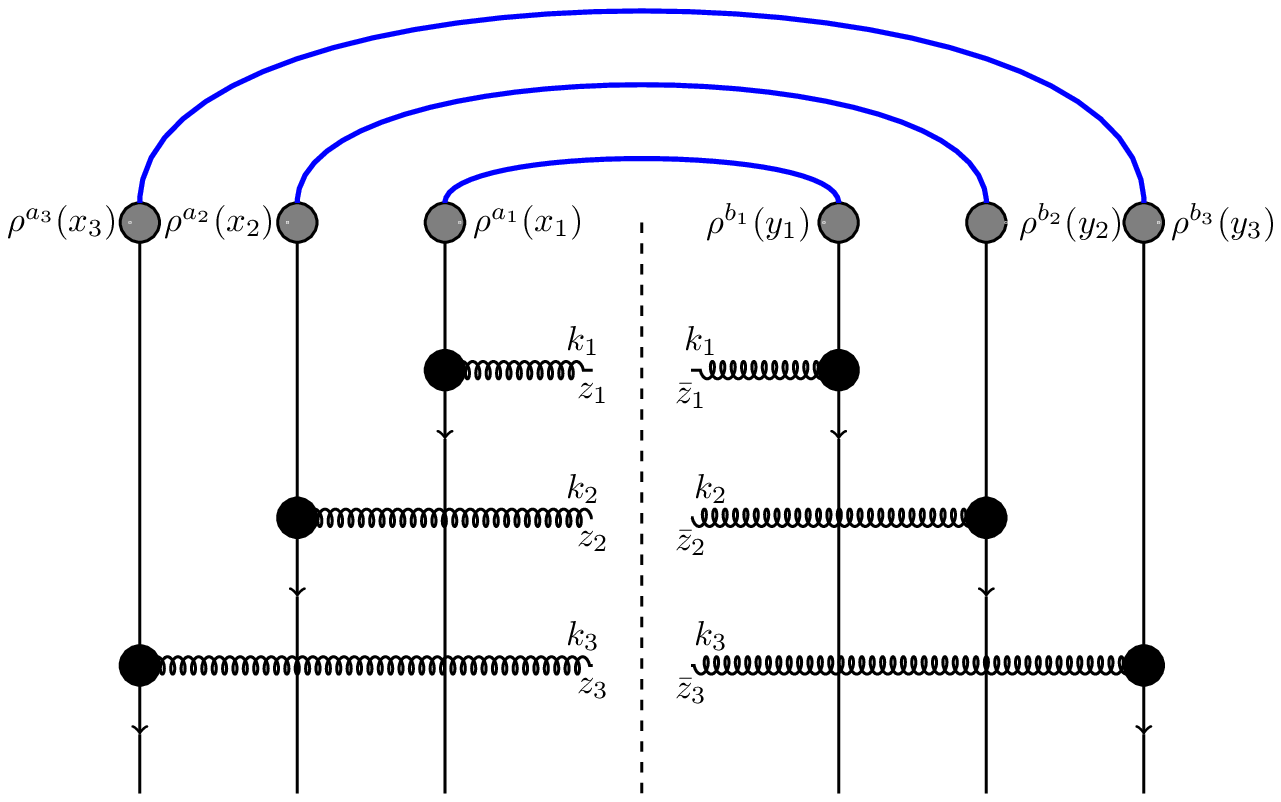}
\end{center}
\caption{Three-dipole (ddd) contribution to the three-gluon production cross section.} 
\label{fig:3d_contribution}
\end{figure}

\begin{figure}[hbt]
\begin{center}
\vspace*{0.5cm}
\includegraphics[width=18cm]{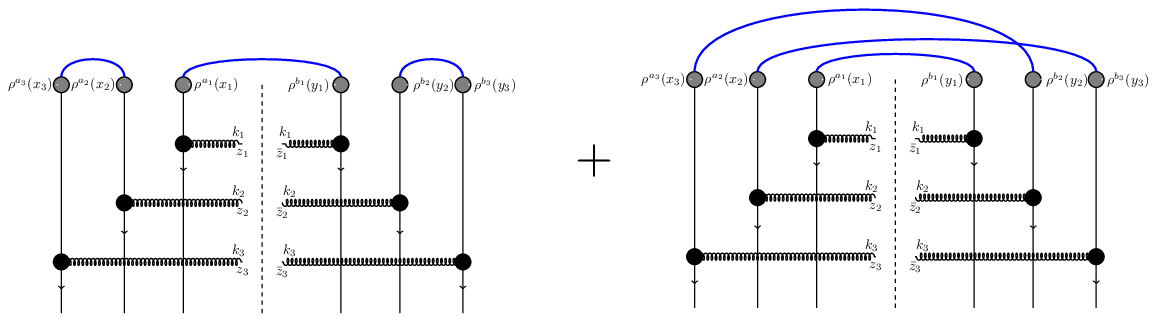}
\end{center}
\caption{Graphical illustration of the first term of the dQ-contribution to the three-gluon production cross section. It corresponds to the independent emission of gluon $k_1$ and interference of gluon $k_2$ and gluon $k_3$.} 
\label{fig:dQ_cont_1st_term}
\end{figure}

\begin{figure}
 [hbt]
\begin{center}
\vspace*{0.5cm}
\includegraphics[width=18cm]{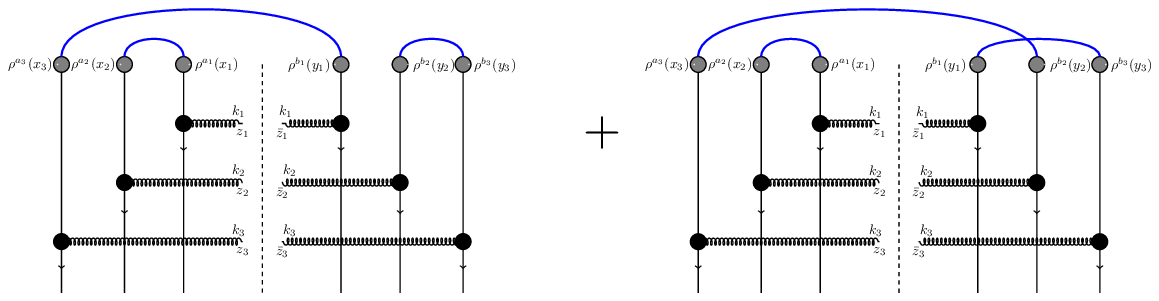}
\end{center}
\caption{Graphical illustration of the term that is proportional to $\big\langle \rho^{a_1}_{x_1} \rho^{a_2}_{x_2}  \big\rangle$ in the sextupole contribution.} 
\label{fig:X_cont_1st_term}
\end{figure}

These terms can be categorized in three main groups. We have labeled the first term as three-dipole (ddd) contribution  since it leads to the product of  three dipoles when multiplied by the target scattering matrices  $U$. The graphical illustration of this term is presented in Fig. \ref{fig:3d_contribution}. The next three terms are named as  dipole-quadrupole (dQ) contribution since they generate a dipole and a quadrupole term after multiplication by $U$'s. One of these terms is illustrated in Fig. \ref{fig:dQ_cont_1st_term}. The last four terms in Eq. \eqref{proj_contraction} are labeled as sextupole (X) contribution since these terms generate the trace of all six Wilson lines. One of these terms (the one that is proportional to $\big\langle \rho^{a_1}_{x_1} \rho^{a_2}_{x_2}  \big\rangle$) is illustrated in Fig. \ref{fig:X_cont_1st_term}. 


We start with the three-dipole contribution to the three-gluon production cross section. We use Eq. \eqref{rho_rho} and substitute the first term of Eq. \eqref{proj_contraction} in the three-gluon production cross section. A straightforward algebra gives the three-dipole contribution as 
\beq
\label{3d_cont_to_3g_production}
&&
\frac{d\sigma}{d^2k_1d\eta_1 \, d^2k_2d\eta_2 \, d^2k_3d\eta_3}\bigg|_{\rm ddd}=  \alpha_s^3 \, (4\pi)^3 \, 
\int_{z_1\zb_1z_2\zb_2z_3\zb_3} e^{ik_1\cdot(z_1-\zb_1)+ik_2\cdot(z_2-\zb_2)+ik_3\cdot(z_3-\zb_3)}
\int _{x_1y_1 x_2y_2 x_3y_3} 
 \nonumber\\
&&
\times 
A^{i}(x_1-z_1)A^{i}(\zb_1-y_1)
 A^{j}(x_2-z_2)A^{j}(\zb_2-y_2)
\, A^{k}(x_3-z_3)A^{k}(\zb_3-y_3)\;  \mu^2(x_1,y_1)\mu^2(x_2,y_2)\mu^2(x_3,y_3)
\nonumber\\
&&
\times
\bigg\langle 
\tr\big\{ [U_{z_1}-U_{x_1}] [U^\dagger_{\zb_1}-U^\dagger_{y_1}]\big\} \; 
\tr \big\{ [U_{z_2}-U_{x_2}][ U^\dagger_{\zb_2}-U^\dagger_{y_2}]\big\} \; 
\tr \big\{ [U_{z_3}-U_{x_3} ] [ U^\dagger_{\zb_3}-U^\dagger_{y_3} ] \big\}
\bigg\rangle_T\ .
\eeq 

We can now Fourier transform the three-dipole contribution. Using the standard definition of the dipole amplitude, Eq. \eqref{dipole_def},
we can write the three-dipole contribution to the three-gluon production as 
\beq
\label{3d_cont_to_3g_production_final}
\frac{d\sigma}{d^2k_1d\eta_1 \, d^2k_2d\eta_2 \, d^2k_3d\eta_3}\bigg|_{\rm ddd} &= &\, \alpha_s^3 \, (4\pi)^3 \, (N_c^2-1)^3 
\int \frac{d^2q_1}{(2\pi)^2} \frac{d^2q_2}{(2\pi)^2} \frac{d^2q_3}{(2\pi)^2}\frac{d^2q_4}{(2\pi)^2} \frac{d^2q_5}{(2\pi)^2} \frac{d^2q_6}{(2\pi)^2} \\
&&\times \big\langle D(q_1,q_2)D(q_3,q_4)D(q_5,q_6) \big\rangle_T\nonumber
\\
&&
\times \; 
\mu^2\big[ k_1-q_1,q_2-k_1\big]  \, \mu^2\big[ k_2-q_3,q_4-k_2\big]  \, \mu^2\big[k_3-q_5,q_6-k_3\big] \; \nonumber\\
&&
\times \; L^i(k_1,q_1)L^i(k_1,q_2) \, L^j(k_2,q_3)L^j(k_2,q_4)  \,  L^k(k_3,q_5)L^k(k_3,q_6),
\nonumber
\eeq
where the function $L^i$, defined in Eq. \eqref{L_def}, gives the transverse momentum structure.
\\

Next, we consider the dipole-quadrupole (dQ) contribution to the three-gluon production cross section. The second, third and  fourth terms in Eq. \eqref{proj_contraction} fall into this category and, using these three terms, we can write the dQ-contribution as 
\beq
\label{dQ_cont_to_3g_production}
&&
\frac{d\sigma}{d^2k_1d\eta_1 \, d^2k_2d\eta_2 \, d^2k_3d\eta_3}\bigg|_{\rm dQ}=  \alpha_s^3 \, (4\pi)^3 \, 
\int_{z_1\zb_1z_2\zb_2z_3\zb_3} e^{ik_1\cdot(z_1-\zb_1)+ik_2\cdot(z_2-\zb_2)+ik_3\cdot(z_3-\zb_3)}
\int _{x_1y_1 x_2y_2 x_3y_3} 
 \nonumber\\
&&
\times 
A^{i}(x_1-z_1)A^{i}(\zb_1-y_1)
 A^{j}(x_2-z_2)A^{j}(\zb_2-y_2)
\, A^{k}(x_3-z_3)A^{k}(\zb_3-y_3) \; \\
&&
\times
\bigg\langle
\mu^2(x_1,y_1)\tr\big\{ [U_{z_1}-U_{x_1}][U^\dagger_{\bz_1}-U^\dagger_{y_1}]\big\} 
\big\lgroup \mu^2(x_2,x_3)\mu^2(y_2,y_3)\, 
\tr\big\{ [U_{\bz_2}-U_{y_2}][U^\dagger_{z_2}-U^\dagger_{x_2}] [U_{z_3}-U_{x_3}][U^\dagger_{\bz_3}-U^\dagger_{y_3}]\big\}\nonumber\\
&&
\hspace{5.5cm}
+\mu^2(x_2,y_3)\mu^2(x_3,y_2) \, \tr\big\{[U_{z_2}-U_{x_2}][U^\dagger_{\bz_2}-U^\dagger_{y_2}][U_{z_3}-U_{x_3}][U^\dagger_{\bz_3}-U^\dagger_{y_3}]\big\}\big\rgroup
\nonumber\\
&&
+\, 
\mu^2(x_2,y_2)\tr\big\{ [U_{z_2}-U_{x_2}][U^\dagger_{\bz_2}-U^\dagger_{y_2}]\big\} 
\big\lgroup
\mu^2(x_1,x_3)\mu^2(y_1,y_3) \, \tr\big\{ [U_{\bz_1}-U_{y_1}][U^\dagger_{z_1}-U^\dagger_{x_1}][U_{z_3}-U_{x_3}][U^\dagger_{\bz_3}-U^\dagger_{y_3}]\big\}
\nonumber\\
&&
\hspace{5.5cm}
+
\mu^2(x_1,y_3)\mu^2(x_3,y_1) \, \tr\big\{ [U_{z_1}-U_{x_1}][U^\dagger_{\bz_1}-U^\dagger_{y_1}] [U_{z_3}-U_{x_3}][U^\dagger_{\bz_3}-U^\dagger_{y_3}]\big\}
\big\rgroup\nonumber\\
&&
+
\mu^2(x_3,y_3) \tr \big\{ [U_{z_3}-U_{x_3}] [U^\dagger_{\bz_3}-U^\dagger_{y_3}]\big\}
\big\lgroup
\mu^2(x_1,x_2)\mu^2(y_1,y_2) \, \tr\big\{ [U_{\bz_1}-U_{y_1}][U^\dagger_{z_1}-U^\dagger_{x_1}][U_{z_2}-U_{x_2}][U^\dagger_{\bz_2}-U^\dagger_{y_2}]\big\}
\nonumber\\
&&
\hspace{5.5cm}
+
\mu^2(x_1,y_2)\mu^2(x_2,y_1) \, \tr\big\{ [U_{z_1}-U_{x_1}][U^\dagger_{\bz_1}-U^\dagger_{y_1}] [U_{z_2}-U_{x_2}][U^\dagger_{\bz_2}-U^\dagger_{y_2}]\big\}
\big\rgroup \bigg\rangle_T\ .\nonumber
\eeq

The graphical illustration of the first term in Eq. \eqref{dQ_cont_to_3g_production} is shown in Fig. \ref{fig:dQ_cont_1st_term}. This term represents the independent emission of the gluon $k_1$ while gluons $k_2$ and $k_3$ interfere. Indeed, the interference of the gluons $k_2$ and $k_3$ are exactly the type A  and type C contributions introduced in the double inclusive gluon production calculation. The remaining two terms in Eq. \eqref{dQ_cont_to_3g_production} correspond to independent emission of the gluon $k_2$ with interference of gluons $k_1$ and $k_3$, and independent emission of the gluon $k_3$ with interference of gluons $k_1$ and $k_2$, respectively.

Fourier transform of Eq. \eqref{dQ_cont_to_3g_production} can be performed in the same way as before. Then, one can use the dipole (Eq. \eqref{Trans_Inv_dipole}) and quadrupole (Eq. \eqref{Trans_Inv_Q}) amplitudes in momentum space and write the dipole-quadrupole contribution to the three-gluon production cross section as 
\beq
\label{dQ_cont_to_3g_production_final}
&&
\frac{d\sigma}{d^2k_1d\eta_1 \, d^2k_2d\eta_2 \, d^2k_3d\eta_3}\bigg|_{\rm dQ}=  \alpha_s^3 \, (4\pi)^3 \, (N_c^2-1)^2
\int \frac{d^2q_1}{(2\pi)^2} \frac{d^2q_2}{(2\pi)^2} \frac{d^2q_3}{(2\pi)^2} \frac{d^2q_4}{(2\pi)^2} \frac{d^2q_5}{(2\pi)^2} \frac{d^2q_6}{(2\pi)^2}
\nonumber\\
&&
\times
\bigg\{ \Big\langle D(q_1,q_2) Q(q_3,q_4,q_5,q_6)\Big\rangle_T\  {\cal L}^{\rm dQ}\big(k_2,q_3,q_4;k_1,q_1,q_2; k_3,q_5,q_6\big) \nonumber \\
&& \hskip 0.5cm +(k_1,q_1,q_2)\leftrightarrow(k_2,q_3,q_4)\nonumber \\
&& \hskip 0.5cm +(k_2,q_3,q_4)\leftrightarrow (k_3,q_5,q_6)\bigg\},
\eeq 
where the function ${\cal L}^{\rm dQ}$ is defined as
\beq
\label{function_cal_L}
&&
{\cal L}^{\rm dQ}\big(k_2,q_3,q_4;k_1,q_1,q_2; k_3,q_5,q_6\big)= 
\mu^2\big[k_2 - q_3, q_4- k_2)\big] L^j(k_2,q_3)L^j(k_2,q_4) 
\nonumber \\
&&
\hspace{-0.3cm}
\times
\Big\{ \mu^2\big[ k_1+q_2, k_3-q_5\big] \; \mu^2\big[ -(k_1-q_1), q_6-k_3)\big]  L^i(k_1,-q_1)L^i(k_1,-q_2)\nonumber \\
&& \hskip 0.5cm +\mu^2\big[k_1-q_1,q_6-k_3\big] \mu^2\big[k_3-q_5,q_2-k_1\big] L^i(k_1,q_1)L^i(k_1,q_2)\Big\} L^k(k_3,q_5)L^k(k_3,q_6)\,.
\eeq

Finally, let us consider the sextupole (X) contribution to the three-gluon production cross section. This contribution stems from the last four terms in Eq. \eqref{proj_contraction}. These are terms that include interference of  all the three gluons . We have shown the illustration of the term that is proportional to $\big\langle \rho^{a_1}_{x_1} \rho^{a_2}_{x_2}  \big\rangle$ in Fig. \ref{fig:X_cont_1st_term}. After contracting the color indexes of these four terms with the Wilson line structure from the target side, we can write the X-contribution to the three-gluon production cross section as 
\beq
\label{X_cont_to_3g_production}
&&
\frac{d\sigma}{d^2k_1d\eta_1 \, d^2k_2d\eta_2 \, d^2k_3d\eta_3}\bigg|_{\rm X}=  \alpha_s^3 \, (4\pi)^3 \, 
\int_{z_1\zb_1z_2\zb_2z_3\zb_3} e^{ik_1\cdot(z_1-\zb_1)+ik_2\cdot(z_2-\zb_2)+ik_3\cdot(z_3-\zb_3)}
\int _{x_1y_1 x_2y_2 x_3y_3} 
 \\
&&
\times 
A^{i}(x_1-z_1)A^{i}(\zb_1-y_1)
 A^{j}(x_2-z_2)A^{j}(\zb_2-y_2)
\, A^{k}(x_3-z_3)A^{k}(\zb_3-y_3) \; 
\nonumber\\
&&
\times \; 
\bigg\langle \mu^2(x_2,x_3)\Big\lgroup \mu^2(x_1,y_2)\, \mu^2(y_1,y_3) \, 
\tr \big\{ [ U_{\zb_1}-U_{y_1} ]  [ U^\dagger_{z_1}-U^\dagger_{x_1} ]  [ U_{\bz_2}-U_{y_2} ]  [U^\dagger_{z_2}-U^\dagger_{x_2}] [U_{z_3}-U_{x_3}] [U^{\dagger}_{\bz_3}-U^\dagger_{y_3}] \big\} 
\nonumber\\
&&
\hspace{2.cm}
+\, 
\mu^2(x_1,y_3)\mu^2(y_1,y_2) \,
\tr\big\{ [U_{z_1}-U_{x_1}] [U^\dagger_{\bz_1}-U^\dagger_{y_1} ] [U_{\bz_2}-U_{y_2}] [U^\dagger_{z_2}-U_{x_2}]  [ U_{z_3}-U_{x_3} ] [U^\dagger_{\bz_3}-U^\dagger_{y_3}]\big\} \Big\rgroup
\nonumber\\
&&
\hspace{0.05cm}
+
\; 
\mu^2(x_2,x_1)\Big\lgroup \mu^2(x_3,y_2) \mu^2(y_3,y_1) \; 
\tr\big\{ [U_{\bz_3}-U_{y_3}]  [U^\dagger_{z_3}-U^\dagger_{x_3}]  [U_{\bz_2}-U_{y_2}] [U^\dagger_{z_2}-U^\dagger_{x_2}] [U_{z_1}-U_{x_1}] [U^\dagger_{\bz_1}-U^\dagger_{y_1}]\big\}
\nonumber\\
&&
\hspace{2.cm}
+\, 
\mu^2(x_3,y_1)\mu^2(y_3,y_2) \, 
\tr\big\{ [U_{z_3}-U_{x_3}] [U^\dagger_{\bz_3}-U^\dagger_{y_3} ]  [U_{\bz_2}-U_{y_2}]  [U^\dagger_{z_2}-U^\dagger_{x_2}] [U_{z_1}-U_{x_1}] [U^\dagger_{\bz_1}-U^\dagger_{y_1}]\big\}\Big\rgroup\nonumber\\
&&
\hspace{0.05cm}
+\; 
\mu^2(x_2,y_1) \Big\lgroup \mu^2(x_1,x_3)\mu^2(y_2,y_3) \; 
\tr\big\{ [U_{z_1}-U_{x_1}] [U^\dagger_{\bz_1}-U^\dagger_{y_1}] [U_{z_2}-U_{x_2}] [U^\dagger_{\bz_2}-U^\dagger_{y_2}] [U_{\bz_3}-U_{y_3}] [U^\dagger_{z_3}-U^\dagger_{x_3}]\big\}
\nonumber\\
&&
\hspace{2.cm}
+\, 
\mu^2(x_1,y_3)\mu^2(x_3,y_2) \; 
\tr\big\{ [U_{z_1}-U_{x_1}] [U^\dagger_{\bz_1}-U^\dagger_{y_1}] [U_{z_2}-U_{x_2}] [U^\dagger_{\bz_2}-U^\dagger_{y_2}] [U_{z_3}-U_{x_3}] [U^\dagger_{\bz_3}-U^\dagger_{y_3}]\big\}\Big\rgroup
\nonumber\\
&&
\hspace{0.05cm}
+\; 
\mu^2(x_2,y_3) \Big\lgroup \mu^2(x_3,x_1)\mu^2(y_2,y_1) \; 
\tr\big\{ [U_{z_3}-U_{x_3}] [U^\dagger_{\bz_3}-U^\dagger_{y_3}] [U_{z_2}-U_{x_2}] [U^\dagger_{\bz_2}-U^\dagger_{y_2}] [U_{\bz_1}-U_{y_1}] [U^\dagger_{z_1}-U^\dagger_{x_1}]\big\}
\nonumber\\
&&
\hspace{2.cm}
+\, 
\mu^2(x_3,y_1)\mu^2(x_1,y_2) \; 
\tr\big\{ [U_{z_3}-U_{x_3}] [U^\dagger_{\bz_3}-U^\dagger_{y_3}] [U_{z_2}-U_{x_2}] [U^\dagger_{\bz_2}-U^\dagger_{y_2}] [U_{z_1}-U_{x_1}] [U^\dagger_{\bz_1}-U^\dagger_{y_1}]\big\}\Big\rgroup\bigg\rangle_T\, .
\nonumber
\eeq

The sextuple amplitude is defined in the usual way as
\beq
{\rm X}(x_1,x'_1,x_2,x'_2,x_3,x'_3)=\frac{1}{N_c^2-1}\tr\left[U({x_1})U^\dagger({x'_1})U({x_2})U^\dagger({x'_2})U({x_3})U^\dagger({x'_3})\right]
\eeq
and its momentum space expression can be written as 
\beq
&& \label{sextupole_momentum}
{\rm X}(x_1,x'_1,x_2,x'_2,x_3,x'_3)=\\
&& \int \frac{d^2q_1}{(2\pi)^2} \frac{d^2q_2}{(2\pi)^2} \frac{d^2q_3}{(2\pi)^2} \frac{d^2q_4}{(2\pi)^2} \frac{d^2q_5}{(2\pi)^2} \frac{d^2q_6}{(2\pi)^2}
\; e^{-iq_1\cdot x_1+iq_2\cdot x'_1-iq_3\cdot x_2 +iq_4\cdot x'_2-iq_5\cdot x_3+i q_6 \cdot x'_3}\, {\rm X}(q_1,q_2,q_3,q_4,q_5,q_6)\,.\nonumber
\eeq
Finally, we can write the sextupole contribution the three-gluon production as 
\beq
\label{X_cont_to_3g_production_final}
&&
\hspace{-0.cm}
\frac{d\sigma}{d^2k_1d\eta_1 \, d^2k_2d\eta_2 \, d^2k_3d\eta_3}\bigg|_{\rm X}=  \alpha_s^3 \, (4\pi)^3 \, (N_c^2-1)
\int \frac{d^2q_1}{(2\pi)^2} \frac{d^2q_2}{(2\pi)^2} \frac{d^2q_3}{(2\pi)^2} \frac{d^2q_4}{(2\pi)^2} \frac{d^2q_5}{(2\pi)^2} \frac{d^2q_6}{(2\pi)^2}
\\
&&
\hspace{-0.cm}
\times 
\bigg\{ \Big\langle {\rm X}(q_1,q_2,q_3,q_4,q_5,q_6)\Big\rangle_T \bigg[ {\cal L}^{\rm X}_1\big(k_1,q_1,q_2;k_2,q_3,q_4;k_3,q_5,q_6\big)+ {\cal L}^{\rm X}_2\big(k_1,q_1,q_2;k_2,q_3,q_4;k_3,q_5,q_6\big)\bigg]\nonumber \\
&&
\hspace{0.6cm}
+(k_1,q_1,q_2)\leftrightarrow (k_3,q_5,q_6) \bigg\},\nonumber
\eeq
where
\beq
\label{function_L_1}
&&
{\cal L}^{\rm X}_1\big(k_1,q_1,q_2;k_2,q_3,q_4;k_3,q_5,q_6\big)
=
\mu^2\big[ k_2-q_3, q_2-k_1\big] L^i(k_1,q_1) L^i(k_1,q_2) L^j(k_2,q_3) L^j(k_2,q_4)
\\
&&
\times
\bigg\{ 
\mu^2\Big[ k_1-q_1,k_3+q_6 \Big]
\mu^2\Big[ q_4-k_2,-k_3-q_5\Big]  L^k(k_3,-q_5) L^k(k_3, -q_6)
\nonumber\\
&&
\hspace{0.5cm}
+ \mu^2\Big[k_1-q_1,q_6-k_3\Big] \mu^2 \Big[k_3-q_5,q_4-k_2\Big] L^k(k_3,q_5) L^k(k_3,q_6)\bigg\}
\nonumber
 \eeq
 and 
 \beq
\label{function_L_2}
&&
{\cal L}^{\rm X}_2\big(k_1,q_1,q_2;k_2,q_3,q_4;k_3,q_5,q_6\big)
=
\mu^2\big[ k_2+q_6, k_1-q_1\big] L^i(k_1,q_1) L^i(k_1,q_2) L^j(k_2,-q_5) L^j(k_2,-q_6)
\\
&&
\times
\bigg\{ 
\mu^2\Big[ k_3+q_4,-k_2-q_5 \Big]
\mu^2\Big[ -k_3-q_3,q_2-k_1\Big]  L^k(k_3,-q_3) L^k(k_3, -q_4)
\nonumber\\
&&
\hspace{0.5cm}
+ \mu^2\Big[k_3-q_3,q_2-k_1\Big] \mu^2 \Big[q_4-k_3,-k_2-q_5\Big] L^k(k_3,q_3) L^k(k_3,q_4)\bigg\}\,.
\nonumber
 \eeq
\subsection{Target averaging in triple inclusive gluon production }
\label{tripletarave}
In this Subsection we perform the target averaging for the three-gluon production cross section by adopting the same procedure introduced in Subsection \ref{doubletarave}. We use Eq. \eqref{Average_2W} for the product of two Wilson lines and calculate the pairwise factorized expressions for the three-dipole, dipole-quadrupole and sextuple contributions to the triple inclusive gluon production cross section. In order to identify the ``irreducible'' quantum interference effects that involve all three gluons, we need to calculate the  triple inclusive gluon production cross section to ${\cal O}[(N_c^2-1)^{-2}]$. Therefore, we will present the results for the pairwise contraction of three-dipoles, dipole-quadrupole and sextuple amplitudes to all orders in number of colors but we will only take into account the relevant terms when calculating the explicit expressions for each contribution. We will assume translational invariance for the dipole, Eq. \eqref{Trans_Inv_dipole2}, when computing the cross sections.

Let us start with the three-dipole contribution. The result for the pairwise factorization of a generic three-dipole amplitude to all orders in $N_c$ reads
\beq
\label{3_dipole_Target_Ave}
&&
\Big\langle D(x_1,x'_1) D(x_2,x'_2) D(x_3,x'_3)\Big\rangle_T\longrightarrow d(x_1,x'_1)d(x_2,x'_2)d(x_3,x'_3)
\\
&&
+\frac{1}{(N_c^2-1)^2}\bigg\{ 
d(x_1,x'_1)\Big[ d(x_2,x_3)d(x'_2,x'_3)+d(x_2,x'_3)d(x'_2,x_3)\Big]
+
d(x_2,x'_2)\Big[ d(x_1,x_3)d(x'_1,x'_3)+d(x_1,x'_3)d(x'_1,x_3)\Big]
\nonumber\\
&&
\hspace{1.4cm}
+\; d(x_3,x'_3)\Big[ d(x_1,x_2)d(x'_1,x'_2)+d(x_1,x'_2)d(x'_1,x_2)\Big]\bigg\}
\nonumber\\
&&
+
\frac{1}{(N_c^2-1)^4}\bigg\{ 
d(x_1,x_2)\Big[ d(x_3,x'_1)d(x'_2,x'_3)+d(x'_1,x'_3)d(x_3,x'_2)\Big]
+
d(x_1,x'_2)\Big[ d(x_3,x'_1)d(x_2,x'_3)+d(x'_1,x'_3)d(x_3,x_2)\Big]
\nonumber\\
&&
\hspace{1.4cm}
+\; 
d(x_1,x_3)\Big[ d(x_2,x'_1)d(x'_3,x'_2)+d(x'_1,x'_2)d(x_2,x'_3)\Big]
+
d(x_1,x'_3)\Big[ d(x_2,x'_1)d(x_3,x'_2)+d(x'_1,x'_2)d(x_2,x_3)\Big]
\bigg\}.
\nonumber
\eeq 

Substituting this factorized expression
into Eq. \eqref{3d_cont_to_3g_production} we  write the three-dipole contribution to the triple inclusive gluon production cross section as
\beq
\label{ddd_Ave_Simplified_integrated}
&&
\frac{d\sigma}{d^2k_1d\eta_1 \, d^2k_2d\eta_2 \, d^2k_3d\eta_3}\bigg|_{\rm ddd}=  \alpha_s^3 \, (4\pi)^3 \, (N_c^2-1)^3
\int \frac{d^2q_1}{(2\pi)^2} \frac{d^2q_2}{(2\pi)^2} \frac{d^2q_3}{(2\pi)^2} \; d(q_1)d(q_2)d(q_3)
\\
&&
\times 
\bigg\{ I_{\rm ddd,0}
+\frac{1}{(N_c^2-1)^2} 
 \Big[ I_{\rm ddd,1}+I_{\rm ddd,2}+I_{\rm ddd,3}\Big]
 +{\cal O}\left(\frac{1}{(N_c^2-1)^4}\right)\bigg\},\nonumber
\eeq
where we have defined  $I_{\rm ddd,0}$ as 
\beq
&&
 I_{\rm ddd,0}= \mu^2(k_1-q_1,q_1-k_1) \, \mu^2(k_2-q_2,q_2-k_2) \, \mu^2(k_3-q_3,q_3-k_3) \; L^i(k_1,q_1)L^i(k_1,q_1)\, L^j(k_2,q_2)L^j(k_2,q_2)
\nonumber\\
 &&
 \hspace{1cm}
 \times\; L^k(k_3,q_3)L^k(k_3,q_3)\; .
\eeq
Moreover, for the ${\cal O}\left(\frac{1}{(N_c^2-1)^{2}}\right)$ terms we have introduced a compact notation 
\beq
I_{\rm ddd,1}=\tilde{I}_{\rm ddd,1}+(k_3\to-k_3),
\eeq
with
 \beq
 &&
 \tilde{I}_{\rm ddd,1}= \mu^2(k_1-q_1,q_1-k_1) \, \mu^2(k_2-q_2,q_3-k_2) \, \mu^2(k_3-q_3,q_2-k_3) \; L^i(k_1,q_1)L^i(k_1,q_1)\, L^j(k_2,q_2)L^j(k_2,q_3)\nonumber\\
 &&
  \hspace{1cm}
 \times\; L^k(k_3,q_3)L^k(k_3,q_2).
\eeq
The remaining terms can be defined by using the explicit expression of $I_{\rm ddd,1}$ and the symmetry properties: 
\beq
&&
I_{\rm ddd,2}\equiv \tilde{I}_{\rm ddd,1}(1\leftrightarrow 2)+(k_3\to-k_3),
\\
&&
I_{\rm ddd,3}\equiv \tilde{I}_{\rm ddd,1}(1\leftrightarrow 3)+(k_2\to-k_2).
\eeq


The pairwise contraction of a generic dipole-quadrupole term can be written as 
\beq
\label{dQ_Target_Ave}
&&
\hspace{-0.6cm} \Big\langle D(x_1,x'_1)Q(x_2,x'_2,x_3,x'_3)\Big\rangle_T\longrightarrow d(x_1,x'_1)\Big[ d(x_2,x'_2)d(x_3,x'_3)+d(x_2,x'_3)d(x_3,x'_2) \Big]
\\
&&\hspace{-0.3cm} +\frac{1}{N_c^2-1}d(x_1,x'_1)d(x_2,x_3)d(x'_2,x'_3)
\nonumber\\
&&
\hspace{-0.3cm} +
\frac{1}{(N_c^2-1)^2}\bigg\{
d(x_1,x_2)\Big[ d(x'_1,x'_2)d(x_3,x'_3)+d(x'_1,x'_3)d(x'_2,x_3)\Big]+d(x_1,x'_2)\Big[d(x'_1,x_2)d(x_3,x'_3)+d(x'_1,x_3)d(x_2,x'_3)\Big]
\nonumber\\
&&
\hspace{0.6cm}
+
d(x_1,x_3)\Big[ d(x'_1,x'_3)d(x_2,x'_2)+d(x'_1,x'_2)d(x'_3,x_2)\Big]+d(x_1,x'_3)\Big[d(x'_1,x_3)d(x_2,x'_2)+d(x'_1,x_2)d(x_3,x'_2)\Big]\bigg\}
\nonumber \\
&&
\hspace{-0.3cm} +
\frac{1}{(N_c^2-1)^3}
\bigg\{ 
d(x'_2,x'_3)\Big[ d(x_1,x_2)d(x'_1,x_3)+d(x_1,x_3)d(x'_1,x_2)\Big] + d(x_2,x_3)\Big[ d(x_1,x'_2)d(x'_1,x'_3)+d(x'_1,x'_2)d(x_1,x'_3)\Big]\bigg\}.
\nonumber
\eeq
Using Eq. \eqref{dQ_Target_Ave} in Eq. \eqref{dQ_cont_to_3g_production}, one can organize this contribution to the triple inclusive gluon production cross section as 
\beq
\label{dQ_Ave_Simplified_integrated}
&&
\frac{d\sigma}{d^2k_1d\eta_1 \, d^2k_2d\eta_2 \, d^2k_3d\eta_3}\bigg|_{\rm dQ}=   \alpha_s^3 \, (4\pi)^3 \, (N_c^2-1)^2
\int \frac{d^2q_1}{(2\pi)^2} \frac{d^2q_2}{(2\pi)^2} \frac{d^2q_3}{(2\pi)^2} \; d(q_1)d(q_2)d(q_3) \nonumber \\
&&
\times
\bigg\{
\Big[ I_{\rm dQ,1}+ I_{\rm dQ,2}+I_{\rm dQ,3}\Big]
+\frac{1}{N_c^2-1}
\Big[ I'_{\rm dQ,1}+I'_{\rm dQ,2}+I'_{\rm dQ,3}\Big]
+{\cal O}\left( \frac{1}{(N_c^2-1)^2}\right)
\bigg\},
\eeq
where we have introduced the same notation used in the three-dipole contribution and we define
\beq
&&
I_{\rm dQ,1}=\tilde{I}_{\rm dQ,1}+(k_2\to-k_2),
\label{def_dQ_1}
\\
&&
I'_{\rm dQ,1}=\tilde{I}'_{\rm dQ,1}+(k_2\to-k_2),
\label{def_dQ_1'}
\eeq
with
\beq
&&
\tilde{I}_{\rm dQ,1}=\mu^2(k_1-q_1,q_1-k_1)\, \mu^2(k_2-q_2,q_3-k_3)\, \mu^2(k_3-q_3,q_2-k_2) \; L^i(k_1,q_1)L^i(k_1,q_1) \, L^j(k_2,q_2)L^j(k_2,q_2) 
\nonumber\\
&&
\hspace{1.2cm}
\times \;  
L^k(k_3,q_3)L^k(k_3,q_3)
\nonumber
\\
&&
\hspace{0.9cm}
+\, 
\mu^2(k_1-q_1,q_1-k_1)\, \mu^2(k_2-q_2,q_2-k_3)\, \mu^2(k_3-q_3,q_3-k_2) \; L^i(k_1,q_1)L^i(k_1,q_1) \, L^j(k_2,q_2)L^j(k_2,q_3) 
\nonumber\\
&&
\hspace{1.2cm}
\times \;  
L^k(k_3,q_3)L^k(k_3,q_2),
\label{def_dQ_1p}
\\
&&
\tilde{I}'_{\rm dQ,1}= \mu^2(k_1-q_1,q_1-k_1)\, \mu^2(-k_2-q_3,q_2+k_3)\, \mu^2(k_2-q_2,q_3-k_3) \; L^i(k_1,q_1)L^i(k_1,q_1) \, L^j(k_2,q_2)L^j(k_2,-q_3) 
\nonumber\\
&&
\hspace{1.2cm}
\times \;  
L^k(k_3,q_3)L^k(k_3,-q_2).
\label{def_dQ_1'p}
\eeq
The remaining terms  can again be written by using the symmetry properties and they are defined as 
\beq
&& 
I_{\rm dQ,2}= \tilde{I}_{\rm dQ,1}(1\leftrightarrow 2)+(k_1\to-k_1) \; , \hspace{2cm}   I_{\rm dQ,3}= \tilde{I}_{\rm dQ,1}(1\leftrightarrow 3)+(k_1\to-k_1), \\
&&
I'_{\rm dQ,2}= \tilde{I}'_{\rm dQ,1}(1\leftrightarrow 2)+(k_1\to-k_1) \; , \hspace{2cm}   I'_{\rm dQ,3}= \tilde{I}'_{\rm dQ,1}(1\leftrightarrow 3)+(k_1\to-k_1) .
\eeq

In a similar manner, one can calculate the pairwise contraction of a generic sextupole term:
\beq
\label{X_Target_Ave}
&&
\Big\langle {\rm X}(x_1,x'_1,x_2,x'_2,x_3,x'_3)\Big\rangle_T\longrightarrow
d(x_1,x'_1)d(x_2,x'_2)d(x_3,x'_3) + d(x_1,x'_3)d(x_2,x'_1)d(x_3,x'_2)
%
\nonumber\\
&&
\hspace{4.12cm}
+\; 
d(x_1,x'_1)d(x_2,x'_3)d(x_3,x'_2) + d(x_2,x'_2)d(x_3,x'_1)d(x_1,x'_3) + d(x_3,x'_3)d(x_1,x'_2)d(x_2,x'_1)
\nonumber\\
&&
+
\frac{1}{N_c^2-1}
\bigg\{ 
d(x_1,x'_1)d(x_2,x_3)d(x'_2,x'_3) + d(x_2,x'_2)d(x_3,x_1)d(x'_3,x'_1) + d(x_3,x'_3)d(x_1,x_2)d(x'_1,x'_2)
\nonumber\\
&& 
\hspace{1.4cm}
+\; 
d(x_2,x_3)d(x_1,x'_3)d(x'_1,x'_2) + d(x_3,x_1)d(x_2,x'_1)d(x'_2,x'_3) + d(x_1,x_2)d(x_3,x'_2)d(x'_3,x'_1)
\bigg\}
\nonumber\\
&&
+
\frac{1}{(N_c^2-1)^2}
\bigg\{
d(x_1,x_2)d(x_3,x'_1)d(x'_2,x'_3) + d(x_2,x_3)d(x_1,x'_2)d(x'_3,x'_1) + d(x_3,x_1)d(x_2,x'_3)d(x'_1,x'_2)
\nonumber\\
&&
\hspace{2cm}
+
d(x_1,x'_2)d(x_3,x'_1)d(x_2,x'_3)
 \bigg\}.
\eeq
We can now substitute Eq. \eqref{X_Target_Ave} into the sextupole contribution to the triple inclusive gluon production cross section given by Eq. \eqref{X_cont_to_3g_production}. The result reads

\beq
\label{dQ_Ave_Simplified_integrated2}
&&
\frac{d\sigma}{d^2k_1d\eta_1 \, d^2k_2d\eta_2 \, d^2k_3d\eta_3}\bigg|_{\rm X}=   \alpha_s^3 \, (4\pi)^3 \, {(N_c^2-1)}
\int \frac{d^2q_1}{(2\pi)^2} \frac{d^2q_2}{(2\pi)^2} \frac{d^2q_3}{(2\pi)^2} \; d(q_1)d(q_2)d(q_3)
 \nonumber\\
&&
\times
\bigg\{
\Big[ I_{\rm X,1}+ I_{\rm X,2}+I_{\rm X,3}+I_{\rm X,4}+I_{\rm X,5} \Big]+ {\cal O}\left( \frac{1}{(N_c^2-1)}\right)+ {\cal O}\left( \frac{1}{(N_c^2-1)^2}\right)\bigg\},
\eeq
where
\beq
I_{\rm X,1}=\Big[\tilde{I}_{\rm X,1}+(k_3\to-k_3)\Big] +\Big[\tilde{I}'_{\rm X,1}+(k_1\to-k_1)\Big]
\eeq
with $\tilde{I}_{\rm X,1}$ and $\tilde{I}'_{\rm X,1}$  defined as 
\beq
&&
\tilde{I}_{\rm X,1}=\mu^2(k_2-q_2, q_2-k_1) \, \mu^2(k_1-q_1, q_3-k_3) \, \mu^2(k_3-q_3, q_1-k_2) \; L^i(k_1,q_1)L^i(k_1,q_2) \, L^j(k_2,q_2)L^j(k_2,q_1) 
\nonumber\\
&&
\hspace{1.5cm}
\times\,  L^k(k_3,q_3)L^k(k_3,q_3)
\nonumber\\
&&
\hspace{1cm}
+\; 
\mu^2(k_2+q_2, k_1-q_2) \, \mu^2(k_3-q_3, q_1-k_1) \, \mu^2(q_3-k_3, -q_1-k_2) \; L^i(k_1,q_1)L^i(k_1,q_2) \, L^j(k_2,-q_1)L^i(k_2,-q_2) 
\nonumber\\
&&
\hspace{1.5cm}
\times\,  L^k(k_3,q_3)L^k(k_3,q_3),
\\
&&
\tilde{I}'_{\rm X,1}=\mu^2(k_1-q_2, q_2-k_2) \, \mu^2(k_2-q_1, k_3-q_3) \, \mu^2(q_1-k_1, q_3-k_3) \; L^i(k_1,q_1)L^i(k_1,q_2) \, L^j(k_2,q_1)L^j(k_2,q_2) 
\nonumber\\
&&
\hspace{1.5cm}
\times\,  L^k(k_3,q_3)L^k(k_3,q_3)
\nonumber\\
&&
\hspace{1cm}
+\; 
\mu^2(-k_1-q_2, q_2-k_2) \, \mu^2(k_2+q_1, q_3-k_3) \, \mu^2(k_3-q_3, k_1-q_1) \; L^i(k_1,-q_2)L^i(k_1,q_1) \, L^j(k_2,-q_1)L^j(k_2,q_2) 
\nonumber\\
&&
\hspace{1.5cm}
\times\,  L^k(k_3,q_3)L^k(k_3,q_3).
\eeq
The terms $I_{\rm X,2}$ and $I_{\rm X,3}$ can again be defined by using the symmetry properties as 
\beq
&&
I_{\rm X,2}=\Big[\tilde{I}_{\rm X,1}(1\to2,2\to3, 3\to1)+(k_3\to-k_3)\Big] +\Big[\tilde{I}'_{\rm X,1}(1\to2,2\to3, 3\to1)+(k_1\to-k_1)\Big],\\
&&
I_{\rm X,3}=\Big[\tilde{I}_{\rm X,1}(1\to3,3\to2, 2\to1)+(k_3\to-k_3)\Big] +\Big[\tilde{I}'_{\rm X,1}(1\to3,3\to2, 2\to1)+(k_1\to-k_1)\Big].
\eeq
The explicit expressions for the remaining two terms read
\beq
&&
I_{\rm [X,4]}=
\mu^2(k_2-q_2, q_1-k_1) \, \mu^2(k_1-q_1, q_3-k_3) \, \mu^2(k_3-q_3, q_2-k_2)  \; L^i(k_1,q_1) L^i(k_1,q_1) \, L^j(k_2,q_2) L^j(k_2,q_2)
\nonumber\\
&&
\hspace{1.5cm}
\times\; 
L^k(k_3,q_3) L^k(k_3,q_3) + (k_3\to-k_3)
\nonumber\\
&&
\hspace{1cm}
+\, 
\mu^2(k_2-q_2, q_3-k_3) \, \mu^2(k_3-q_3, k_1-q_1) \, \mu^2(q_1-k_1, q_2-k_2)  \; L^i(k_1,q_1) L^i(k_1,q_1) \, L^j(k_2,q_2) L^j(k_2,q_2)
\nonumber\\
&&
\hspace{1.5cm}
\times\; 
L^k(k_3,q_3) L^k(k_3,q_3) + (k_1\to-k_1)
\nonumber\\
&&
\hspace{1cm}
+\, 
\mu^2(k_2-q_2, k_1-q_1) \, \mu^2(k_3-q_3, q_1-k_1) \, \mu^2(q_3-k_3, q_2-k_2)  \; L^i(k_1,q_1) L^i(k_1,q_1) \, L^j(k_2,q_2) L^j(k_2,q_2)
\nonumber\\
&&
\hspace{1.5cm}
\times\; 
L^k(k_3,q_3) L^k(k_3,q_3) + (k_3\to-k_3)
\nonumber\\
&&
\hspace{1cm}
+\, 
\mu^2(q_1-k_1, q_3-k_3) \, \mu^2(k_1-q_1, q_2-k_2) \, \mu^2(k_2-q_2, k_3-q_3)  \; L^i(k_1,q_1) L^i(k_1,q_1) \, L^j(k_2,q_2) L^j(k_2,q_2)
\nonumber\\
&&
\hspace{1.5cm}
\times\; 
L^k(k_3,q_3) L^k(k_3,q_3) + (k_1\to-k_1),
\eeq
\beq
&&
I_{\rm [X,5]}=
\mu^2(k_2-q_2, q_2-k_1) \, \mu^2(k_1-q_1, q_1-k_3) \, \mu^2(k_3-q_3, q_3-k_2)  \; L^i(k_1,q_1) L^i(k_1,q_2) \, L^j(k_2,q_2) L^j(k_2,q_3)
\nonumber\\
&&
\hspace{1.5cm}
\times\; 
L^k(k_3,q_3) L^k(k_3,q_1) + (k_3\to-k_3)
\nonumber\\
&&
\hspace{1cm}
+\, 
\mu^2(k_2-q_2, q_2-k_3) \, \mu^2(k_3-q_3, q_3+k_1) \, \mu^2(q_1-k_2, -k_1-q_1)  \; L^i(k_1,-q_1) L^i(k_1,-q_3) \, L^j(k_2,q_2) L^j(k_2,q_1)
\nonumber\\
&&
\hspace{1.5cm}
\times\; 
L^k(k_3,q_3) L^k(k_3,q_2) + (k_1\to-k_1)
\nonumber\\
&&
\hspace{1cm}
+\, 
\mu^2(k_2+q_1, k_1-q_1) \, \mu^2(k_3-q_3, q_3-k_1) \, \mu^2(q_2-k_3, -k_2-q_2)  \; L^i(k_1,q_1) L^i(k_1,q_3) \, L^j(k_2,-q_2) L^j(k_2,-q_1)
\nonumber\\
&&
\hspace{1.5cm}
\times\; 
L^k(k_3,q_3) L^k(k_3,q_2) + (k_3\to-k_3)
\nonumber\\
&&
\hspace{1cm}
+\, 
\mu^2(k_2+q_3, k_3-q_3) \, \mu^2(-k_1-q_1, q_1-k_3) \, \mu^2(k_1-q_2, q_2-k_2)  \; L^i(k_1,-q_1) L^i(k_1,q_2) \, L^j(k_2,q_2) L^j(k_2,-q_3)
\nonumber\\
&&
\hspace{1.5cm}
\times\; 
L^k(k_3,q_3) L^k(k_3,q_1) + (k_1\to-k_1).
\eeq


Finally, the triple inclusive gluon production cross section can be organized according to the powers in the number of colors and the result reads 
\beq
\label{3g_full_cross_section}
&&
\frac{d\sigma}{d^2k_1d\eta_1 \, d^2k_2d\eta_2 \, d^2k_3d\eta_3}=   \alpha_s^3 \, (4\pi)^3 \, (N_c^2-1)^3
\int \frac{d^2q_1}{(2\pi)^2} \frac{d^2q_2}{(2\pi)^2} \frac{d^2q_3}{(2\pi)^2} \; d(q_1)d(q_2)d(q_3) 
\nonumber\\
&&
\times
\bigg\{  I_{\rm ddd,0}
+\frac{1}{N_c^2-1}
\Big[ I_{\rm dQ,1}+ I_{\rm dQ,2} + I_{\rm dQ,3} \Big]
\nonumber\\
&&
+\frac{1}{(N_c^2-1)^2}
\bigg\lgroup
\Big[ I_{\rm ddd,1}+I_{\rm ddd,2}+I_{\rm ddd,3}\Big] + \Big[ I'_{\rm dQ,1}+I'_{\rm dQ,2}+I'_{\rm dQ,3}\Big]
+\Big[ I_{\rm X,1}+I_{\rm X,2}+I_{\rm X,3}+I_{\rm X,4}+I_{\rm X,5}\Big]
\bigg\rgroup
\nonumber\\
&&
+{\cal O}\left(\frac{1}{(N_c^2-1)^3}\right)+{\cal O}\left(\frac{1}{(N_c^2-1)^4}\right)\bigg\}\; .
\label{tripleres}
\eeq

This is our final result for the triple inclusive gluon production cross section explicitly written up to order $(N_c^2-1)^{-3}$. It is straightforward to calculate the $(N_c^2-1)^{-3}$ and $(N_c^2-1)^{-4}$ terms with all the ingredients introduced in this Subsection. However, as mentioned earlier, in order to observe the quantum interference effects it is enough to calculate the triple inclusive gluon production cross section to order $(N_c^2-1)^{-2}$. Thus, we have not written the higher order terms explicitly. 

\subsection{Identifying terms in triple inclusive gluon production }
\label{tripleid}
Now, we can take a closer look at each term in the triple inclusive gluon production cross section separately and identify them one by one. We will follow the same logic that was introduced in Subsection \ref{doubleiden} to perform this analysis, i.e., we will use the fact that 
\beq
\mu^2(k,p)\propto F[(k+p)R],
\eeq
where function $F$ is a soft form factor that is peaked around zero and $R$ is the radius of the projectile. 
\\

\noindent (i) {\it  ${\cal O}(1)$ terms}: 

\begin{itemize}
\item  The only contribution that we get at ${\cal O}(1)$ is the $I_{\rm ddd,0}$ term. It is clear that this term is the classical contribution to triple inclusive gluon production cross section and it is equal to the  product of three single inclusive gluon production cross sections.  It is responsible for independent emission of   all three gluons, with transverse momenta $k_1$, $k_2$ and $k_3$, and does not generate any correlations. 

\end{itemize}

\noindent (ii) {\it  ${\cal O}(1/N_c^{2})$ terms}: 

At this order, we have three different terms: $I_{\rm dQ,1}$, $I_{\rm dQ,2}$ and $I_{\rm dQ,3}$. Let us start with $I_{\rm dQ,1}$ whose explicit expression can be read off from  Eqs. \eqref{def_dQ_1}, \eqref{def_dQ_1'}, \eqref{def_dQ_1p} and \eqref{def_dQ_1'p}.
%
%
\begin{itemize}
\item $I_{\rm dQ,1}$ term has two contributions:

The first one is proportional to  
\beq
\mu^2(k_1-q_1,q_1-k_1)\, \mu^2(k_2-q_2,q_3-k_3)\, \mu^2(k_3-q_3,q_2-k_2)
&\propto& \mu^2(k_1-q_1,q_1-k_1) \nonumber\\
&&
\times\; 
F^2\left\{\left[ (k_2-q_2)-(k_3-q_3)\right]R\right\}.
\eeq
The form factor is peaked around $k_2-q_2=k_3-q_3$ while the gluon $k_1$ does not interfere with the remaining two gluons. Thus, it is clear that this term is a contribution to forward peak of Bose enhancement of gluons $k_2-q_2$ and $k_3-q_3$ in the projectile with the third gluon emitted independently. The mirror image of this term, given by the transformation $k_2\to-k_2$, contributes to the backward peak of the Bose enhancement of the gluons $k_2-q_2$ and $k_3-q_3$ in the projectile. Obviously, in this case the third gluon is emitted independently as well.  

The second term of $I_{\rm dQ,1}$ is proportional to
\beq
\mu^2(k_1-q_1,q_1-k_1)\, \mu^2(k_2-q_2,q_2-k_3)\, \mu^2(k_3-q_3,q_3-k_2) \propto \mu^2(k_1-q_1,q_1-k_1)F^2\left[(k_2-k_3)R\right].
\eeq
In this contribution the form factor is peaked around $k_2=k_3$. Therefore, it is clear that this term is a contribution to the forward peak of the HBT correlations of the gluons $k_2$ and $k_3$ while the third gluon is emitted independently. The mirror image of this term is given by the transformation $k_2\to-k_2$ and will be a contribution to the backward peak of the HBT correlations of the gluons $k_2$ and $k_3$.

Since $I_{\rm dQ,2}$ and $I_{\rm dQ,3}$ are related to $I_{\rm dQ,1}$ by symmetry, it is obvious that these terms exhibit the same behavior  as $I_{\rm dQ,1}$ but with gluons interchanged ($1\leftrightarrow 2$ and $1\leftrightarrow 3$ respectively). 

\end{itemize}

\noindent (iii) {\it  ${\cal O}(1/N^4_c)$ terms}:

\begin{itemize}
\item $I_{\rm ddd,1}$ term: 

This term is proportional to 
\beq
\mu^2(k_1-q_1,q_1-k_1) \, \mu^2(k_2-q_2,q_3-k_2) \, \mu^2(k_3-q_3,q_2-k_3) \propto \mu^2(k_1-q_1,q_1-k_1) F^2\left[(q_3-q_2)R\right].
\eeq
In this term, the form factor is peaked around  $q_2=q_3$ where $q_2$ and $q_3$ are the momenta of the gluons in the target wave function. Therefore, this term is clearly a contribution to the forward peak of the Bose enhancement of the gluons $q_2$ and $q_3$ in the target wave function while the third gluon is emitted independently.  Its mirror image given by the transformation $k_3\to-k_3$ is a contribution to the backward peak of the Bose enhancement of the gluons $q_2$ and $q_3$ in the target wave function.
 
The remaining two terms that stem from the three-dipole contribution at ${\cal O}(1/N^4_c)$ are $I_{\rm ddd,2}$ and $I_{\rm ddd,3}$. These terms can be obtained by exchanging ($1\leftrightarrow 2$) and ($1\leftrightarrow 3$). Hence, they exhibit the same behavior as  $I_{\rm ddd,1}$. Namely,  $I_{\rm ddd,2}$ is a contribution to the (forward/backward peaks) Bose enhancement of the gluons $q_1$ and $q_3$ in the target wave function while gluon $q_2$ is emitted independently, and $I_{\rm ddd,3}$ is a contribution to the (forward/backward peaks of the) Bose enhancement of the gluons $q_1$ and $q_2$ in the target wave function while the gluon $q_3$ is emitted independently.

\item $I'_{\rm dQ,1}$ term:

This term is proportional to 
\beq
\hspace{-0.5cm}
\mu^2(k_1-q_1,q_1-k_1)\, \mu^2(-k_2-q_3,q_2+k_3)\, \mu^2(k_2-q_2,q_3-k_3) &\propto& \mu^2(k_1-q_1,q_1-k_1) \nonumber\\
&&
\times\; 
F^2\left\{\left[ (k_2-q_2)-(k_3-q_3)\right]R\right\}.
\eeq

The form factor is again peaked around $k_2-q_2=k_3-q_3$. Therefore, this term contributes to the forward peak of the Bose enhancement of the gluons $k_2-q_2$ and $k_3-q_3$ in the projectile wave function with the third gluon emitted independently. Clearly, the mirror image is a contribution to the backward peak.  
$I'_{\rm dQ,2}$ and $I'_{\rm dQ,3}$ exhibit the same behavior with the exchange of ($1\leftrightarrow 2$ ) and ($1\leftrightarrow 3$). 

We would like to emphasize that these three terms are $N_c$-suppressed corrections to the (forward/backward peaks of the) Bose enhancement of the two gluons in the projectile wave function while the third gluon is emitted independently, which was the behavior that we have observed in the first part of the $I_{\rm dQ,1}$, $I_{\rm dQ,2}$ and $I_{\rm dQ,3}$ terms. 

\item $I_{\rm X,1}$ term: 

This is the first term that we are analyzing which stems from the sextupole contribution. Unlike the previous terms that we have analyzed, all the terms that originate from the sextupole contribution lead to the interference of all three gluons and there are no independent emissions. After this short comment, let us take a closer look at $I_{\rm X,1}$ term. It is composed of four different contributions. The first contribution is proportional to 
\beq
&&\mu^2(k_2-q_2, q_2-k_1) \, \mu^2(k_1-q_1, q_3-k_3) \, \mu^2(k_3-q_3, q_1-k_2)\nonumber \\ &&\propto F\left[(k_2-k_1)R\right]F^2\left\{\left[(k_1-q_1)-(k_3-q_3)\right]R\right\}.
\eeq
The first form factor is peaked around $k_1=k_2$ while the other two form factors peaked around $k_1-q_1=k_3-q_3$. Thus, this term is a contribution to the forward peak of the HBT correlations of gluons $k_1$ and $k_2$ together with the forward peak of the Bose enhancement of gluons $k_1-q_1$ and $k_3-q_3$ in the projectile wave function.  Its mirror image, given by the transformation $k_3\to-k_3$, is a contribution to backward peak of the Bose enhancement of $k_1-q_1$ and $k_3-q_3$ in the projectile wave function together with a contribution to the forward peak of the HBT correlations of gluons $k_1$ and $k_2$.

The second term in $I_{\rm X,1}$ is proportional to 
\beq
&&\mu^2(k_2+q_2, k_1-q_2) \, \mu^2(k_3-q_3, q_1-k_1) \, \mu^2(q_3-k_3, -q_1-k_2)\nonumber \\ &&\propto F\left[(k_2+k_1)R\right]F^2\left\{\left[(k_1-q_1)-(k_3-q_3)\right]R\right\}.
\eeq
The first form factor is peaked around $k_1=-k_2$ and the remaining two form factors are the same as the previous contribution. Therefore, one can identify this term as a contribution to the backward peak of the HBT correlations of the gluons $k_1$ and $k_2$ together with the forward peak of the Bose enhancement of the gluons $k_1-q_1$ and $k_3-q_3$ in the projectile wave function.  Obviously, its mirror image, given by the transformation $k_3\to-k_3$, is a contribution to the backward peak of HBT correlations of gluons $k_1$ and $k_2$ together with the contribution to the backward peak of the Bose enhancement of the gluons $k_1-q_1$ and $k_3-q_3$ in the projectile wave function.

The third term in  $I_{\rm X,1}$ is proportional to 
\beq
&&\mu^2(k_1-q_2, q_2-k_2) \, \mu^2(k_2-q_1, k_3-q_3) \, \mu^2(q_1-k_1, q_3-k_3)\nonumber \\ &&\propto F\left[(k_2-k_1)R\right]F^2\left\{\left[-(k_1-q_1)-(k_3-q_3)\right]R\right\}.
\eeq
The first form factor in this term is peaked around $k_1=k_2$ while the other two form factors are peaked around $k_1-q_1=q_3-k_3$. Thus, it is clear that this term is a contribution to the forward peak of HBT correlations of gluons $k_1$ and $k_2$ together with the backward peak of the Bose enhancement of gluons $k_1-q_1$ and $k_3-q_3$ in the projectile. Its mirror image, given by the transformation $k_1\to-k_1$, is a contribution to the backward peak of HBT correlations  of gluons $k_1$ and $k_2$ together with the forward peak of the Bose enhancement of gluons $k_1-q_1$ and $k_3-q_3$.

The last term in $I_{\rm X,1}$ is proportional to 
\beq
&&\mu^2(-k_1-q_2, q_2-k_2) \, \mu^2(k_2+q_1, q_3-k_3) \, \mu^2(k_3-q_3, k_1-q_1)\nonumber \\
&&\propto F\left[(-k_2-k_1)R\right]F^2\left\{\left[(k_1-q_1)+(k_3-q_3)\right]R\right\}.
\eeq
The first form factor is peaked around $k_1=-k_2$ and the remaining two form factors are peaked around $k_1-q_1=q_3-k_3$. Therefore, it is a contribution to the
 backward peak of HBT correlations of gluons $k_1$ and $k_2$ together with the backward peak of the Bose enhancement of gluons $k_1-q_1$ and $k_3-q_3$ in the projectile wave function. Its mirror image, given by the transformation $k_1\to-k_1$, is a contribution to the forward peak of HBT  correlations of gluons $k_1$ and $k_2$ together with the forward peak of the Bose enhancement of gluons $k_1-q_1$ and $k_3-q_3$ in the projectile wave function.
 \\

It can easily be shown that $I_{\rm X,2}$ and $I_{\rm X,3}$ exhibit a similar behavior as $I_{\rm X,1}$ with gluons interchanged. $I_{\rm X,2}$ contributes to (backward/forward) HBT correlations of gluons $k_2$ and $k_3$ together with (backward/forward) Bose enhancement of gluons $k_1-q_1$ and $k_2-q_2$ in the projectile wave function. Finally,  $I_{\rm X,3}$ contributes to (backward/forward) HBT correlations of gluons $k_1$ and $k_3$ together with (backward/forward) Bose enhancement of gluons $k_2-q_2$ and $k_3-q_3$ in the projectile wave function.

\item $I_{\rm X,4}$ term: 

This term has four different contributions and they all contribute to the Bose enhancement of all three gluons. 

The first contribution in $I_{\rm X,4}$ is proportional to 
\beq
&&
\hspace{-2cm}
\mu^2(k_2-q_2, q_1-k_1) \, \mu^2(k_1-q_1, q_3-k_3) \, \mu^2(k_3-q_3, q_2-k_2)\propto F\{[(k_2 -q_2)-(k_1 -q_1)]R\}  \nonumber\\
&&
\hspace{2cm}
\times\,F\{[(k_1 -q_1)-(k_3 -q_3)]R\} \,  F\{[(k_3 -q_3)-(k_2 -q_2)]R\} .
\eeq
The first form factor is peaked around $k_1-q_1=k_2-q_2$, the second one is peaked around $k_1-q_1=k_3-q_3$ and the last one is peaked around $k_2-q_2=k_3-q_3$. Therefore, it is clear that this term is a contribution to the forward peak of the Bose enhancement of gluons $k_1-q_1$ and $k_2-q_2$, gluons $k_1-q_1$ and $k_3-q_3$, and gluons $k_2-q_2$ and $k_3-q_3$ in the projectile wave function. Its mirror image, given by the transformation $k_3\to-k_3$, is a contribution to the forward peak of the Bose enhancement of the gluons $k_2-q_2$ and $k_1-q_1$ together with the backward peak of the Bose enhancement of the gluons $k_1-q_1$ and $k_3-q_3$, and gluons $k_2-q_2$ and $k_3-q_3$ in the projectile wave function. 

The second contribution in $I_{\rm X,4}$ is proportional to 
\beq
&&
\mu^2(k_2-q_2, q_3-k_3) \, \mu^2(k_3-q_3, k_1-q_1) \, \mu^2(q_1-k_1, q_2-k_2)  \propto F\{[(k_2 -q_2)-(k_3 -q_3)]R\}  \nonumber\\
&&
\hspace{2cm}
\times\,F\{[(k_1 -q_1)+(k_3 -q_3)]R\} \,  F\{[(q_1-k_1)+(q_2-k_2)]R\} .
\eeq
Clearly, this is a contribution to the forward peak of Bose enhancement of gluons $k_2-q_2$ and $k_3-q_3$ together with the backward peak of the Bose enhancement of gluons $k_1-q_1$ and $k_2-q_2$, and gluons $k_1-q_1$ and $k_3-q_3$ in the projectile wave function. Its mirror image is given by the transformation $k_1\to-k_1$. Therefore, it contributes to the forward peak of Bose enhancement of gluons $k_2-q_2$ and $k_3-q_3$, gluons $k_1-q_1$ and $k_2-q_2$, and gluons $k_1-q_1$ and $k_3-q_3$ in the projectile wave function.

The third contribution in $I_{\rm X,4}$ is proportional to 
\beq
&&
\mu^2(k_2-q_2, k_1-q_1) \, \mu^2(k_3-q_3, q_1-k_1) \, \mu^2(q_3-k_3, q_2-k_2)   \propto F\{[(k_2 -q_2)+(k_1 -q_1)]R\}  \nonumber\\
&&
\hspace{2cm}
\times\,F\{[(k_3 -q_3)-(k_1 -q_1)]R\} \,  F\{[(q_3-k_3)+(q_2-k_2)]R\} .
\eeq
By looking at the peaks of the form factors, it is straightforward to see that this is a contribution to the forward peak of the Bose enhancement of the gluons $k_1-q_1$ and $k_3-q_3$ together with a backward peak of the Bose enhancement of the gluons $k_1-q_1$ and $k_2-q_2$, and gluons $k_3-q_3$ and $k_2-q_2$ in the projectile wave function. Its mirror image is given by the transformation $k_3\to-k_3$. Therefore, it contributes to the forward peak of the Bose enhancement of the gluons $k_2-q_2$ and $k_3-q_3$ together with a backward peak of the Bose enhancement of the gluons $k_1-q_1$ and $k_3-q_3$, and gluons $k_1-q_1$ and $k_2-q_2$ in the projectile wave function.

The last contribution in $I_{\rm X,4}$ is proportional to 
\beq
&&
\mu^2(q_1-k_1, q_3-k_3) \, \mu^2(k_1-q_1, q_2-k_2) \, \mu^2(k_2-q_2, k_3-q_3) \propto F\{[(q_1 -k_1)+(q_3 -k_3)]R\}  \nonumber\\
&&
\hspace{2cm}
\times\,F\{[(k_1 -q_1)-(k_2 -q_2)]R\} \,  F\{[(k_2-q_2)+(k_3-q_3)]R\} .
\eeq
Clearly, this term is a contribution to the forward peak of the Bose enhancement of gluons $k_1-q_1$ and $k_2-q_2$ together with a backward peak of the Bose enhancement of the gluons $k_1-q_1$ and $k_3-q_3$, and gluons $k_2-q_2$ and $k_3-q_3$ in the projectile wave function. Its mirror image is given by the
transformation $k_1\to-k_1$. Thus, it contributes to the forward peak of the Bose enhancement of gluons $k_1-q_1$ and $k_3-q_3$ together with a backward peak of the Bose enhancement of the gluons $k_1-q_1$ and $k_2-q_2$, and gluons $k_2-q_2$ and $k_3-q_3$ in the projectile wave function

\item $I_{\rm X,5}$ term:

The last term that stems from the sextupole contribution at $O(1/N^4_c)$ is the $I_{\rm X,5}$ term. It has four different contributions and all of them contribute to the HBT correlations of all three gluons.

The first contribution in $I_{\rm X,5}$ is proportional to 
\beq
\mu^2(k_2-q_2, q_2-k_1) \, \mu^2(k_1-q_1, q_1-k_3) \, \mu^2(k_3-q_3, q_3-k_2)  \propto F[(k_2-k_1)R] \, F[(k_1-k_3)R]\, F[(k_3-k_2)R] .
\eeq
In this term, the form factors are peaked around $k_2=k_1$, $k_1=k_3$ and $k_3=k_2$. Thus, it is a contribution to the forward peak of the HBT correlations of gluons $k_1$ and $k_2$, gluons $k_1$ and $k_3$, and gluons $k_2$ and $k_3$. The mirror image of this term is given by the transformation $k_3\to-k_3$ and it contributes to the forward peak of the HBT correlations of $k_1$ and $k_2$ together with backward peak of the HBT correlations of the gluons $k_1$ and $k_3$, and gluons  $k_2$ and $k_3$. 

The second contribution in $I_{\rm X,5}$ is proportional to 
\beq
\mu^2(k_2-q_2, q_2-k_3) \, \mu^2(k_3-q_3, q_3+k_1) \, \mu^2(q_1-k_2, -k_1-q_1) \nonumber \\
\propto F[(k_2-k_3)R] \, F[(k_1+k_3)R]\, F[(-k_2-k_1)R] .
\eeq
The form factors in this term are peaked around $k_2=k_3$, $k_1=-k_2$ and $k_1=-k_3$. Hence, this term contributes to the forward peak of the HBT correlations of gluons $k_2$ and $k_3$ together with the backward peak of the HBT correlations of gluons $k_1$ and $k_2$, and gluons $k_1$ and $k_3$. The mirror image of this term is given by the transformation $k_1\to-k_1$. Therefore, it contributes to the forward peak of the HBT correlations of gluons $k_1$ and $k_3$, gluons $k_2$ and $k_3$, and gluons $k_1$ and $k_2$.

The third contribution in $I_{\rm X,5}$ is proportional to 
\beq
\mu^2(k_2+q_1, k_1-q_1) \, \mu^2(k_3-q_3, q_3-k_1) \, \mu^2(q_2-k_3, -k_2-q_2) \nonumber \\
\propto F[(k_2+k_1)R] \, F[(k_3-k_1)R]\, F[(-k_2-k_3)R] .
\eeq
In this term the form factors are peaked around $k_2=-k_1$, $k_3=k_1$ and $k_2=-k_3$. It contributes to the forward peak of HBT correlations of gluons $k_3$ and $k_1$ together with a backward peak of the HBT correlations of gluons $k_1$ and $k_2$, and gluons $k_2$ and $k_3$. On the other hand, its mirror image (given y the transformation $k_3\to-k_3$) contributes to the forward peak of the HBT correlations of gluons $k_3$ and $k_2$ together with the backward peak of the HBT correlations of gluons $k_1$ and $k_2$, and gluons $k_1$ and $k_3$.

Finally, the last contribution in $I_{\rm X,5}$ is proportional to 
\beq
\mu^2(k_2+q_3, k_3-q_3) \, \mu^2(-k_1-q_1, q_1-k_3) \, \mu^2(k_1-q_2, q_2-k_2)\nonumber \\
 \propto F[(k_2+k_3)R] \, F[(-k_3-k_1)R]\, F[(k_1-k_2)R] .
\eeq
Clearly, this term contributes to the forward peak of the HBT correlations of the gluons $k_1$ and $k_2$ together with the backward peak of the  HBT correlations of the gluons $k_2$ and $k_3$, and gluons $k_3$ and $k_1$. Its mirror image is given by the transformation $k_1\to-k_1$. Therefore, it contributes to the forward peak of the HBT correlations of the gluons $k_1$ and $k_3$ together with the backward peak of the  HBT correlations of the gluons $k_2$ and $k_3$, and gluons $k_1$ and $k_2$.

\end{itemize}

\section{Conclusions}
\label{conclu}
To conclude, we have calculated the inclusive two and three gluon production in p-A collisions at mid rapidity in the CGC formalism. We use the generalized McLerran-Venugopalan model to perform the projectile averaging. This model allows for accounting for the finite transverse  size of the projectile.  

We observe that in the full dilute-dense limit that goes beyond the glasma graphs, the expression for the cross section double inclusive gluon production contains two types of terms: a product of two dipole amplitudes and a quadrupole. The origin of the quadrupole term is the contribution to scattering where the two incoming gluons exchange color while propagating through the target.

We further used simple physical assumptions about the target structure to express the quadrupole average in terms of products of averages of two dipoles. We stress that this factorization is not a result of the large $N_c$ limit and is, in fact, completely unrelated with the large $N_c$ expansion. It  is rather the result of the physical expectation that the color neutralization of the target ensemble happens on transverse scales of order $1/Q_s$. As discussed in \cite{amir}, this approximation does not take into account the ``classical'' correlated term arising from the contributions to transverse integrals where the produced particles have to be close to each other in the incoming projectile wave function. 

The resulting expression for  double inclusive production is quite simple, see Eq. \eqref{doubleres}.  It exhibits very similar terms to those obtained for double inclusive quark production in \cite{amir}.  Just like there, we find that all the quantum interference effects that constitute the genuine multiparticle correlations (as can be extracted e.g. using the cumulant method \cite{cumulants}) at order $1/N_c^2$, given by the $I_1$ term in \eqref{doubleres}, originate from the quadrupole.
 Our results in this part of the paper are consistent with \cite{Altinoluk:2018hcu}.
 We observe two types of quantum interference effects - the Bose enhancement of gluons in the projectile wave function and the Hanbury-Brown-Twiss interference effect. In the case of gluons  (differently from the production of identical quarks) the two effects enhance the production of gluons  which are both collinear and anticollinear in the transverse plane. These same effects have been observed in the glasma graph calculation earlier \cite{kevin,us,yuri}. 
 
  Additionally to \cite{Altinoluk:2018hcu} we identify the Bose enhancement terms in the target wave function. There is an interesting difference in this aspect between the full dilute-dense calculation presented here, and the glasma graphs of \cite{kevin,us}. In the glasma graph calculation, which is based on the dilute-dilute limit, and is therefore symmetric between the projectile and the target, the Bose enhancement in the {\it target} wave function is of the leading order in $1/N_c$, just like the Bose enhancement in the projectile.  In the complete dilute-dense framework utilized in the present paper this effect, although present, is suppressed as $1/N_c^2$ relative to the projectile Bose enhancement effect. 
  
  We have also computed the triple inclusive gluon production in the dilute-dense limit for the first time. Our result is given in Eq. \eqref{tripleres}.  Although the expressions are lengthy, the basic physics is very simple. One observes  terms in the production which involve the interference between two of the gluons, and  independent emission of the third one. Such terms would be subtracted if one would calculate the three particle cumulant rather than write up the three particle inclusive cross section. Additionally, there are genuine three particle correlation terms which appear at order $1/N_c^4$ - these are the $I_{\rm X,i}$ terms in \eqref{tripleres}. These terms originate solely from the sextupole in the cross section. 
 They include interference due to Bose enhancement and HBT contributions of all three particles, as well as ``mixed'' terms where two of the particles are coupled via Bose enhancement, and other two due to  the HBT effect. 
 
 These features are very similar to those observed in three quark production in \cite{amir}. There is one interesting albeit expected difference. In production of three quarks the correction due to interference of all three quarks has an opposite sign to the term where one quark is emitted independently of the other two (which interfere). It therefore has a flavor of ``unitarization correction'' as discussed in \cite{amir}. For gluon production this is not the case. All terms in the cross section are positive, and thus the genuine three gluon interference term enhances the correlations rather than weakening them.
  
Correlations among more than two particles have been studied at RHIC and the LHC, e.g., in the context of the study of properties of the produced medium \cite{Abelev:2008ac,Abelev:2009jv,Ajitanand:2010zz}, for the study of HBT correlations \cite{Abelev:2014pja} or for extraction of the azimuthal asymmetries
\cite{Chatrchyan:2013nka,Khachatryan:2015waa}. Without some simplifications and implementation of models for the proton and nuclei, it is very difficult to provide definite predictions beyond the long range pseudorapidity nature of our correlations. Such study and the computation of several gluon inclusive production that is paved by the present work, with the obvious extension to the four gluon inclusive case, are left for future work.

\acknowledgements
AK thanks Amir Rezaeian and Vladi Skokov for many discussions related to the subject of this work. We also thank Mauricio Mart\'{\i}nez, Matt Sievert and Doug Wertepny for pointing out some mistakes in Eqs.  \eqref{Final_Type_A_Q}, \eqref{Final_Type_B_dd} and \eqref{Final_Type_C_Q}, and \eqref{3d_cont_to_3g_production_final}, \eqref{dQ_cont_to_3g_production_final} and \eqref{X_cont_to_3g_production_final}, in the first version of this paper.
TA and AK express their gratitude to the Departamento de F\'{\i}sica de Part\'{\i}culas at Universidade de Santiago de Compostela, for support and hospitality when part of this work was done. ML thanks the hospitality and support of the CERN theory division. We thank Centro de Ciencias de Benasque Pedro Pascual for warm hospitality during the program {\it Collectivity and correlations in high-energy hadron and nuclear collisions} and the {\it COST workshop on collectivity in small systems}.
The work of TA is supported by Grant No. 2017/26/M/ST2/01074 of the National Science Centre, Poland.
NA was supported by  Ministerio de Ciencia e Innovaci\'on of Spain under projects FPA2014-58293-C2-1-P, FPA2017-83814-P and Unidad de Excelencia Mar\'{\i}a de Maetzu under project MDM-2016-0692,  by Xunta de Galicia (Conseller\'{\i}a de Educaci\'on) within the Strategic Unit AGRUP2015/11, and by FEDER. 
AK was supported by the NSF Nuclear Theory grant 1614640, the Fulbright US scholar program and a CERN scientific associateship. ML was supported
by the  Israeli Science Foundation grant \# 1635/16;  AK and ML  were also supported by the
BSF grants \#2012124 and \#2014707. 
This work has been performed in the framework of COST Action CA15213 "Theory of hot matter and relativistic heavy-ion collisions" (THOR).

\end{document}